\documentclass[a4paper,11pt,hyper]{JHEP3}

\title{Instantons, black holes, and harmonic functions}

\author{Thomas Mohaupt \\
Theoretical Physics Division, Department of Mathematical Sciences,
University of Liverpool, Peach Street, Liverpool L69 7ZL, 
United Kingdom\\
E-Mail: \email{Thomas.Mohaupt@liv.ac.uk}
}
\author{Kirk Waite \\
Theoretical Physics Division, Department of Mathematical Sciences,
University of Liverpool, Peach Street, Liverpool L69 7ZL, 
United Kingdom\\
E-Mail: \email{Kirk.Waite@liv.ac.uk}
}

\abstract{
We find a class of five-dimensional Einstein-Maxwell type
Lagrangians which contains the bosonic Lagrangians of 
vector multiplets as a subclass, and preserves some features
of supersymmetry, namely the existence of multi-centered
black hole solutions and of attractor equations. Solutions
can be expressed in terms of harmonic functions through a
set of algebraic equations. The geometry underlying these
Lagrangians is characterised by the existence of a Hesse 
potential and generalizes the very special real geometry
of vector multiplets. 

Our construction proceeds by first obtaining instanton
solutions for a class of four-dimensional Euclidean sigma models, 
which includes those occuring
for four-dim\-en\-sio\-nal Euclidean $N=2$ vector multiplets
as a subclass. For solutions taking values in a
completely isotropic submanifold of the target space,
we show that the solution can be expressed in terms of
harmonic functions if an integrability condition is
met. This condition can either be solved by imposing
that the solution depends on a single coordinate,
or by imposing that the target space is a para-K\"ahler
manifold which can be obtained from a real Hessian manifold
by a generalized $r$-map. In the latter case one 
obtains multi-centered solutions. Moreover, 
if the integrability condition
is met, the second order equations of motion can always
be reduced to first order equations, which become 
gradient flow equations if the solution is further required 
to depend on one coordinate only.
The dualization of axions into tensor fields
and the lifting of four-dimensional instantons to five-dimensional
solitons are used to motivate the addition of a boundary term to 
the action, which accounts for the instanton action. If the
sigma model is coupled to gravity, and if the Hesse potential is
of a suitable form which we specify, then the four-dimensional 
Euclidean Lagrangian can be lifted consistently to a
five-dimensional Einstein-Maxwell type Lagrangian. 
Instanton solutions lift to extremal black hole solutions, and
the instanton action equals the ADM mass.
}


\usepackage{amsmath,amssymb}
\usepackage{amscd}
\usepackage{bbm}

\usepackage{bbold}

\preprint{LTH 832}
\begin{document}


\section{Introduction and Overview}

\subsection{Introduction}

Stationary solutions of supergravity theories, such as black holes, black 
$p$-branes, gravitational waves and Kaluza-Klein monopoles can often
be presented in terms of harmonic functions, which only depend on
the coordinates transverse to the worldvolume.\footnote{There are many 
excellent reviews on the subject,
including \cite{Stelle}, which discusses the approach we are going to use in 
its Section 9.} This is related to
the existence of multi-centered solutions: if the field equations can 
be reduced to a set of decoupled harmonic equations without assuming 
spherical symmetry, 
then not only single-centered harmonic functions, 
\[
H(r) = h + \frac{q}{r^{D-3}}\;,
\]
but also multi-centered harmonic functions
\[
H(\vec{x}) = h + \sum_{i=1}^N \frac{q_i}{|\vec{x} - \vec{x}_i|^{D-3}} \;.
\]
provide solutions of the field equations.
The existence of stationary multi-centered solutions requires the exact
cancellation
of the forces between the constituents at arbitrary distance, 
the classical examples being
multi-centered extremal black hole solutions like the Majumdar-Papapetrou
solutions of Einstein-Maxwell theory \cite{MajPap}. This cancellation is
often explained by supersymmetry: if the theory allows an embedding
into a supersymmetric theory, then one can look for solutions admitting
Killing spinors, which in turn leads to stationary multi-centered solutions.
The saturation of an extremality bound, which is needed for the
cancellation of forces is then equivalent to the saturation 
of the supersymmetric mass bound (also called the BPS mass bound). 
The extremal Reissner-Nordstr\"om
black hole, and its multi-centered generalizations are the prototypical
examples of such supersymmetric solitons \cite{Gib,GibHul}. In 
supergravity theories with $N\geq 2$ supersymmetry the asymptotic
behaviour of BPS solutions at event horizons is determined by the
charges through the black hole attractor mechanism \cite{FerKalStr,Str,FerKal}, 
which forces
the scalar fields to take fixed point values. The attractor mechanism
and the construction of solutions in terms of harmonic functions
are closely related: from 
the attractor equations (also called stabilisation equations or
fixed point equations), which determine
the asymptotic near-horizon solution one can obtain
the so-called generalized stabilisation 
equations, which allow to express the complete solution algebraically
in terms of harmonic functions
\cite{FerKalStr,FerKal,BCdWKLM,Sabra4d,BLS,Sabra5d,ChaSab,CdWKM}.
In the single-centered case 
the generalized stabilisation equations can be formulated equivalently 
as gradient flow equations for the scalars
as functions of the radial coordinate \cite{FGK,Moore,Denef}. 
The potential driving the flow is the central charge.

While imposing supersymmetry is sufficient to derive the attractor 
mechanism, and to obtain multi-centered solutions,
it is not necessary. 
As already observed in \cite{FGK} the attractor mechanism is a
general feature of extremal black holes in Einstein-Maxwell 
type theories.  More recently, 
single-centered 
non-supersymmetric extremal solutions have been studied extensively
and from various perspectives starting from \cite{GIJT,TriTri,KalSivSor}.
Reviews of this subject can be found in 
\cite{Erice,SpringerLN}.
By imposing that the solution is spherically symmetric in addition to 
stationary, one can reduce the problem of solving the equations of motion
to a one-dimensional problem which only involves the radial coordinate
\cite{FGK}. 
The reduced problem is formally equivalent to the motion of a
particle on a curved target space in presence of a potential, usually
called the black hole potential. One contribution to the potential
depends on the charges and is obtained by eliminating the
gauge fields through their equations of motion. Alternatively,
one can often convert the  gauge fields (or at least the those
components of the gauge fields relevant for the solution) into
scalars. Then the equations of motion take the form of a geodesic
equation (without potential) on an extended scalar manifold which
encodes all relevant degrees of freedom.
The black hole potential receives further contributions 
if the full higher-dimensional solution involves rotation, 
if the gauge fields cannot be expressed in terms of scalars,
and if a cosmological
constant, higher curvature terms, Taub-NUT charge, or other such 
complications are present. We will retrict ourselves to situations which can
be formulated as geodesic motion (without potential) 
on an enlarged scalar manifold. In this set-up one is left with
solving the scalar equations of motions, while 
the Einstein equations themselves result in a constraint,
which can be interpreted as the conservation of the particle's energy.

The standard approach to single-centered solutions is to try
rewriting the second order scalar equations of motion as 
first order gradient flow equations. 
While for BPS solutions the potential driving the
gradient flow is the central charge \cite{FGK}, first order rewritings 
have since then been found for various non-BPS solutions, and
the function driving the flow is referred to as the (`fake'-, `generalized' or
`pseudo'-)superpotential or as the prepotential 
\cite{CerDal,CCDOP,PSVV:08,ADOT}. The problem of finding a first
order gradient flow prescription can be reformulated using the
Hamilton-Jacobi formalism as the problem of finding a 
canonical transformation \cite{ADOT}. 
In many cases the first 
order equations can be interpreted as generalized Killing spinor
equations, by defining a suitable covariant derivative for spinors. 
Similar observation have been made before in the context of 
cosmological solutions and domain walls, and this has motivated
the concept of `fake'- or `pseudo'-supersymmetry \cite{Fake1,Fake2}.

While much is known about the attractor mechanism for non-BPS black holes,
we are not aware of a systematic analysis of the conditions which 
allow multi-centered solutions. From the supersymmetric case one 
is used to the observation that the existence of a first order
rewriting and the reduction of the equations of motion to 
algebraic relations between the scalars and a set of harmonic
functions (i.e. generalized stabilization equations) are closely 
related. In the single-centered case one might regard the
generalized stabilisation equations as `solutions' of 
the gradient flow equation.\footnote{This `solution' is in general
not completely explicit, since generically one cannot find explicit
expression for the scalars in terms of the harmonic functions.}
But if one looks beyond single-centered solutions, it becomes 
clear that the generalized stabilization equations are much more
then mere solutions to the gradient flow equations. In the absence
of spherical symmetry, the gradient flow 
equations are replaced by first order partial differential 
equations, which have not been studied much in the literature.
In contrast the form of the generalized stabilization equations 
remains the same, and non-spherical solutions simply correspond
to a more general choice of harmonic functions, namely 
multi-centered instead of single-centered harmonic functions.
Also note that for BPS solutions the
generalized stabilization equations can be derived directly,
without passing through an intermediate stage of first finding
a first order rewriting and then solving the flow equations \cite{BLS,CdWKM}.
This suggest to develop an approach to non-BPS black holes 
which is not based on first order rewriting and flow equations,
but on generalised stabilisation equations. In other words 
one should try to reduce the second order equations of motion 
to decoupled harmonic equations, without imposing spherical
symmetry. We do not expect that this strategy can work for
general Einstein-Maxwell theories. 
As remarked in \cite{FGK}, it is virtually impossible to obtain
detailed information about the behaviour of extremal
black hole solution away from the horizon and infinity
if the scalar manifold is generic. The special 
geometries of vector multiplet manifolds provide examples
where explicit solutions can be found (in general up to 
algebraic equations). Our strategy will be to start 
with general Einstein-Maxwell type Lagrangians, which include
those of $N=2$ vector multiplets as a subclass, and then
to work out the additional constraints that we need to 
impose on the scalar manifold in order to obtain multi-centered
solutions. Since we are only interested in stationary solutions,
we can perform a dimensional
reduction over time and work with the resulting Euclidean
theory. Dimensional reduction over time is a powerful
solution generating technique, which was (to our knowledge)
first used in \cite{NeuKra:69} and \cite{GibBreMai:88},
and which has recently be used to explore non-BPS
extremal black holes (albeit only for single-centered
solutions)  \cite{CerDal, CCDOP,GNPW:05,GaiLiPadi:08,PSVV:08,ChiGut,ADOT}.
We refer to \cite{Stelle,Bergshoeff:Geodesic,EucIII} for 
reviews of this method. 
The essential part of the reduced Lagrangian is a sigma model,
whose equation of motion is the equation for a harmonic
map from the reduced space-`time' to the scalar target space.
The problem which we will investigate in this paper is to reduce
the non-linear second order partial differential equation 
of a harmonic map to decoupled linear harmonic equations.
As we will see this reduction imposes non-trivial conditions
on the scalar manifold, which define a generalized version 
of the special geometry of $N=2$ vector multiplets, and which
is characterised by the existence of a potential for the metric.

While previous studies have either
investigated supergravity Lagrangians, or Lagrangians with
generic scalar manifolds, we have identified 
an interesting intermediate class of scalar manifolds:
they are much more general as the target spaces of 
supergravity theories, while still allowing to express
the solution in terms of harmonic functions and thus
to obtain multi-centered solutions. This class of scalar
manifolds is much more generic than symmetric spaces. 
For symmetric spaces, powerful methods from the theory
of Lie groups are available, and the construction of 
BPS and non-BPS extremal solutions can be related to
intergrable systems  \cite{GNPW:05,GaiLiPadi:08,PSVV:08,ChiGut,CRTV,ADOT}.
While this class of
models is very interesting, symmetric spaces are not even general
enough to cover the scalar geometries supergravity with
$N\leq 2$ supersymmetry. Thus one is limited to 
models with $N>2$ supersymmetry, or to special $N\leq 2$ models, 
like toroidal and orbifold compactifications, or consistent truncations 
of models
with $N>2$ supersymmetry.

While our analysis could (and ultimately should) be carried out 
in an arbitrary number of dimensions, we will be more specific
and fix the number of dimensions to be five. Since 
our approach is guided by results on $N=2$ vector multiplets,
this is a natural choice, because the so-called very special geometry
of five-dimensional vector multiplets \cite{GST} is the simplest of the
special geometries of $N=2$ supermultiplets.  
From our results it will be clear that there
is a similar story for four-dimensional $N=2$ vector multiplets,
but the five-dimensional case is a more
convenient starting point for technical simplicity.
Thus, we will start with generic five-dimensional
Einstein-Maxwell theories and construct asymptotically flat,
electrically charged, extremal, multi-centered solutions 
by using the associated four-dimensional Euclidean sigma models.
By imposing that the equations of motion reduce to decoupled
harmonic equations, we obtain a constraint on the scalar 
metric which generalizes 
the very special real geometry of five-dimensional
vector multiplets \cite{GST}. There are in fact two relevant 
conditions. An integrability condition for the solution implies 
the existence of a Hesse potential for the scalar
metric, while the consistent lifting of the four-dimensional
Euclidean solution to a solution 
of the five-dimensional Einstein-Maxwell theory requires in addition
that the Hesse potential is the logarithm of a homogeneous function,
which we call the prepotential. Five-dimensional supergravity
corresponds to the special case where this prepotential is 
homogeneous of degree three. When expressed in terms of 
five-dimensional variables, the algebraic relations which 
express the solution in terms of harmonic functions take
the form of the generalized stabilization equations for
five-dimensional vector multiplets \cite{Sabra5d,ChaSab}. 
Therefore the solutions contain
the static (non-rotating) electric multi-centered BPS solutions
of five-dimensional supergravity \cite{Sabra5d,ChaSab} as a subclass. 
We also consider the case where we
lift solutions of a four-dimensional Euclidean sigma model
without coupling to gravity. In this case the Hesse potential
is not constrained, and we obtain solitonic solutions of a five-dimensional
gauge theory coupled to scalars.

The construction of solutions is presented from the reduced,
four-dim\-en\-sio\-nal perspective, i.e. we first construct
solutions of four-dimensional Euclidean sigma models and 
discuss the lifting to five dimensions in a second step. 
The target geometries of the four-dimensional sigma models 
include those of four-di\-men\-sio\-nal
Euclidean vector multiplets, both rigid and local, as special cases. 
Therefore there is some overlap between this paper and work 
on Euclidean special geometry \cite{EucI,EucII,EucIII}. In particular,
the target space geometry of the four-dimensional Euclidean sigma model
is para-complex,\footnote{Para-complex geometry is explained in some detail 
in \cite{EucI,EucIII}. The features relevant for our work will be explained 
in due course.} and the integrability
condition guaranteeing the existence of multi-centered solutions
implies that it is para-K\"ahler. The relation between the
real, Hessian target spaces of five-dimensional sigma models and 
the para-K\"ahler geometry of four-dimensional sigma models 
provides a `para-version' or `temporal version' of the generalized
$r$-map described recently in \cite{AleCor}. To be precise, we 
find two different generalized para-$r$-maps, depending on whether we 
couple the Euclidean sigma model to gravity before lifting, or not. 
In \cite{AleCor} only the case witout gravity was considered.

The solutions of the reduced Euclidean
theory can be interpreted as instantons and are interesting 
in their own right. They are of the same type as the D-instanton solution 
of type-IIB supergravity \cite{GGP}, and the instanton solutions
of $N=2$ hypermultiplets \cite{BGLMM,GutSpa,TheVan,DdVTV,deVVAn,ChiGut},
and they contain the instanton solutions
and $N=2$ vector multiplets \cite{EucIII,MohWai1} as a subclass.
Since the instantons satisfy a Bogomol'nyi bound
and lift to extremal black holes, we refer to them as
extremal instanton solutions. Extremality is equivalent to
satisfying what we call the `extremal instanton ansatz,' 
which restricts the scalars
to vary along totally isotropic submanifolds of the scalar target.
This in turn is equivalent to the vanishing of the energy momentum
tensor, which makes it consistent to solve the reduced Einstein 
equations by taking the four-dimensional reduced metric to be flat.
The  dimensional lifting to five dimensions then 
gives rise to extremal black hole solutions. 

For single-centered 
extremal solutions of supergravity theories the distinction between
BPS solutions and non-BPS solutions manifests itself in the 
form of the potential which drives the gradient flow equations.
For BPS solutions this potential is the central charge, 
while for non-BPS solution it is another function, which
one needs to construct.
In our framework this distinction finds a geometric interpretation
in terms of the para-K\"ahler geometry of the target space
of the Euclidean sigma model, because the extremal instanton ansatz 
comes in two versions. The first version, 
which can be imposed without further constraints on the scalar metric
requires that the scalar fields vary along the eigendistributions
of the para-complex structure. However, if the metric has 
discrete isometries, a generalized version of the extremal 
instanton ansatz is possible,
which allows the scalars to vary along other completely isotropic
submanifolds of the target. This distinction generalizes the
one between BPS and non-BPS extremal solutions in supergravity, 
and also provides a geometric interpretation of the difference
between the two types of extremal instanton solutions.

Besides serving as generating solutions for
higher-dimensional solitons, instanton solutions 
are relevant for computing
instanton corrections to quantum amplitudes and effective actions.  
While this second application is not our main focus in this paper, 
we encounter one notorious problem arising in this context:
if one computes the instanton action by substituting the instanton 
solution into the Euclidean action one obtains zero instead of the
expected non-vanishing finite result.
We review one of the proposed solutions, namely the
dualization of axions into tensor fields \cite{GGP}. In this dual picture the 
Euclidean action
is positive definite and extremal instantons satisfy a Bogomol'nyi bound.
This motivates to add a specific 
boundary term to the original `purely scalar' action, which ensures 
that its evaluation on instanton solutions give the same result
as the dual `scalar-tensor' action.
We show that the instanton
action obtained this way 
agrees with the ADM mass of the black hole obtained by lifting
the solution to five dimensions.
If instead we lift four-dimensional
solutions to five dimensions without coupling to gravity,
we again find that the mass of the resulting soliton is
equal to the instanton action.


\subsection{Overview}

This paper is structured as follows. In Section 2.1 we introduce 
the class of Euclidean sigma models which we will use to generate
solutions.  The scalar target space is required to be 
para-Hermitian\footnote{The relevant concepts from para-complex 
geometry will be explained in Section 2.}
and to have $n$ commuting shift isometries.
In Section 2.2 we show that the Euclidean scalar
equations of motion can be reduced 
to a set of linear harmonic
equations by imposing the extremal instanton ansatz, which corresponds
to restricting the scalar fields to vary along totally isotropic
subspaces of the target space.  The consistency of the solution 
leads to an integrability condition, which has two natural solutions.
Either one restricts the solution to depend on one variable only.
Since we require that solutions approach a vacuum at infinity, this implies
spherical symmetry. While this does not impose conditions on the scalar metric, 
it excludes multi-centered solutions. The second, more interesting
solution of the integrability condition 
requires that the scalar metric has a Hesse potential. 
In this case the target space 
is para-K\"ahler rather than only para-Hermitian.
Since no constraint needs to be imposed on the 
solutions themselves, we obtain multi-centered 
solutions. In Section 
2.3 we define instanton charges, which are the conserved charges
corresponding to the $n$ commuting shift symmetries, which are 
required in order to be able to lift the Euclidean sigma model
to a five-dimensional gauge theory. Then we rederive the extremal
instanton solutions from a different angle. 
By imposing that solutions carry 
finite instanton charge, we can `peel of' one derivative 
from the field equations and reduce them to first order 
equations. As long as one does not impose spherical symmetry 
these are still (quasi-linear) partial differential equations.
But once spherical symmetry is imposed, which we do in Section 2.4,
the field equations reduce to first order gradient flow equations. 
We include some observations and remarks about the relation of 
our approach to the one based on first order rewritings, and
to the Hamilton-Jacobi approach.

In Section 3 we discuss a dual version of Euclidean sigma models,
where the $n$ axionic scalars have been dualized into tensor fields.
In the dual formulation the action of extremal instanton solutions
is finite, positive and satisfies a Bogomol'nyi bound. To be precise,
the finiteness of the action requires a suitable behaviour of the
scalar fields at the centers of the harmonic functions. These conditions
are further analyzed in Section 4.
Instead of working with the dual `scalar-tensor' action, one 
can add a boundary to the original `purely scalar' action, which
has the effect that one obtains the same finite
non-vanishing instanton action for both actions.

In Section 4 we analyze two classes of Hesse potentials in more detail:
homogeneous functions and logarithms of homogeneous functions. In 
these cases the asymptotic behaviour of the scalars at the centers 
and at infinity can be determined even if the field equations cannot
be solved in closed form. The conditions which guarantee the finiteness
of the instanton action are found explicitly for this
class: the Hesse potential must be homogeneous of negative degree,
or it must be the logarithm of a homogeneous function (of any degree).
The instanton action can be expressed as a function of the instanton charges
and of the asymptotic scalar fields, which has the standard form
of a BPS mass formula.
We can also find analogues of the stabilization equations and
generalized stabilization equations known from BPS black holes. 
Solutions do not quite show fixed point behaviour, but the scalars
run off to points at infinite affine parameter, with fixed finite
ratios that are determined by the charges. We give various explicit
examples of solutions, which include both rigidly and locally
supersymmetric models as well as models which cannot have a 
supersymmetric extensions (the generic case). 

In Section 5 we briefly discuss the lifting of four-dimensional
Euclidean sigma models to five-dimensional field theories without
gravity. The most interesting result is that the mass of the
resulting soliton equals the instanton action. Since instanton
charges are electric charges from the five-dimensional point
of view, the expression for the mass takes the same form 
as for the BPS mass in a supersymmetric theory.
The special case of
a cubic Hesse potential gives us the rigid para-$r$-map between the
scalar geometries of five-dimensional vector multiplets and 
four-dimensional Euclidean vector multiplets. For general Hesse 
potentials we obtain a `para-version' of the generalized rigid
$r$-map
which relates Hessian manifolds to para-K\"ahler manifolds 
with $n$ commuting shift isometries.

In Section 6.1 we discuss the relation between four-dimensional 
Euclidean sigma models coupled to gravity, 
and five-dimensional Einstein-Maxwell type
theories.
We start in five dimensions, and present a generalized version of the
very special real geometry of vector multiplets where the 
prepotential is allowed to be homogeneous of arbitrary degree.
This is used 
to write down a class of Einstein-Maxwell type Lagrangians, 
which reduce over time to para-K\"ahler sigma models with
$n$ commuting shift isometries, coupled to gravity.
This provides a generalized version of the local 
para-$r$-map, which includes the para-$r$-map between
supersymmetric theories as a special case. We also
indicate how the reduction over space results in a generalized
version of the local $r$-map.

We then set up an instanton -- black hole dictionary.
Lifting extremal instanton solutions gives extremal black holes,
and the ADM mass is shown to be equal to the instanton action
in Section 6.2. In Section 6.3 we turn to the 
entropy of the black holes, which is non-vanishing 
or zero, depending
on how many charges are switched on. The black hole entropy can be
interpreted in the instanton picture 
by using a specific conformal frame for the four-dimensional metric,
which is different from the Einstein frame. We call this frame
the Kaluza-Klein frame, because it corresponds to a 
fixed time slice of the five-dimensional metric. In this frame
the four-dimensional metric of extremal instantons is not
flat, but only conformally flat, and the geometry can be
interpreted as a semi-infinite wormhole. The Bekenstein-Hawking
entropy of the black hole corresponds to the asymptotic size
of the throat of the wormhole, 
and the degenerate 
case of black holes with vanishing entropy
corresponds to wormholes with vanishing asymptotic size of 
the throat.

In Section 6.4 we illustrate the relation between extremal
instantons and extremal black holes with several explicit
examples. Then we show in full generality 
that the instanton attractor equations lift to black hole
attractor equations, which have the same form as the
stabilization equations and generalized stabilization equations
of five-dimensional vector multiplets. In particular,
we show that the `fixed-ratio run-away' behaviour of four-dimensional
scalar is equivalent to the proper fixed point behaviour 
of five-dimensional scalars.

In Appendix A we expand on the observation that target 
space geometries, which are obtained from a higher-dimensional
theory by dimensional reduction 
over space or time, respectively, can be viewed as different
real sections of one underlying complex target space. 
We explain the notion of  `complexifying (para-)complex
numbers' and indicate that complex-Riemannian geometry is the
appropriate framework for relating target spaces
occuring in dimensional reduction over space and time by analytical
continuation. 


\section{Sigma models with para-Hermitean target spaces}

\subsection{Motivation and discussion of the Euclidean
 action \label{SectEucAct}}

The starting point for all subsequent constructions are
sigma models of the form
\begin{equation}
\label{paraH-real}
S[\sigma,b]_{(0,4)} = \int
d^4 x \; \frac{1}{2} N_{IJ}(\sigma) \left( \partial_m \sigma^I
\partial^m \sigma^J - \partial_m b^I \partial^m b^J \right) \;.
\end{equation}
Space-time is taken to be flat Euclidean space $E$
with indices $m=1,2,3,4$. 
The target space $M$ is $2n$-dimensional
with coordinates $\sigma^I, b^I$, where $I=1,\ldots, n$.
The matrix $N_{IJ}(\sigma)$ is assumed to be real, positive 
definite and  only depends on half
of the scalar fields. Thus the metric of $M$ has $n$ commuting 
isometries which act as shifts on the axionic scalars $b^I$:
\begin{equation}
b^I \rightarrow b^I + C^I \;,
\end{equation}
where $C^I$ are constants. 
The relative minus sign between the
kinetic terms of the scalars $\sigma^I$ and the axionic scalars
$b^I$ implies that the metric $N_{IJ} \oplus (-N_{IJ})$ of $M$ has split 
signature $(n,n)$. There are two related reasons for considering Euclidean
sigma models with split signature target spaces:
\begin{enumerate}
\item
One approach to the definition of Euclidean actions combines the
standard Wick rotation with an analytic continuation $b^I \rightarrow
i b^I$ for axionic scalars \cite{vNWal}. D-instantons and other instanton 
solutions
of string theory and supergravity 
are obtained as classical solutions of Euclidean 
actions of this type \cite{GGP,BGLMM,GutSpa,TheVan,DdVTV,ChiGut}. 
\item
Solitons, i.e. stationary, regular, finite energy solutions of 
$(n+1)$-dimensional theories can be dimensionally reduced over time,
resulting in instantons, i.e. regular finite action solutions 
of the reduced $n$-dimensional Euclidean theory. Conversely, one 
approach to the construction of solitons is to reduce the theory
under consideration over time to obtain a simpler Euclidean 
theory, preferably a scalar sigma model. Instanton solutions 
of the Euclidean theory can then be lifted to solitons of the
original theory. String theory has a large variety of solitonic
solutions, which play a central role in establishing string 
dualities and thus obtaining information about the non-perturbative
completion of the theory. Dimensional reduction has been used
as a solution generating technique for some time in Einstein-Maxwell theory
\cite{NeuKra:69}, supergravity \cite{GibBreMai:88} and
string theory \cite{CleGal}. More recent applications include
\cite{GNPW:05,GaiLiPadi:08,PSVV:08,EucIII}. We refer to 
\cite{Stelle,Bergshoeff:Geodesic,EucIII} for a review of this method. 
\end{enumerate}

If we do not couple the sigma model (\ref{paraH-real}) to gravity, then 
its lift to $1+4$ dimensions is\footnote{Dimensional lifiting
in the presence of gravity will be discussed later. As we will see
the results obtained without coupling to gravity remain valid, 
provided that suitable restrictions on the scalar metric are imposed.} 
\begin{equation}
\label{5dAction}
S[\sigma, A, \ldots]_{(1,4)} 
= \int d^5 x \left( - \frac{1}{2} N_{IJ}(\sigma) \partial_\mu \sigma^I
\partial^\mu \sigma^J - \frac{1}{4} N_{IJ}(\sigma) F_{\mu \nu}^I 
F^{J|\mu \nu} + \cdots \right) \;.
\end{equation}
Here space-time is five-dimensional Minkowski space with
indices $\mu, \nu, \ldots = 0,1,2,3,4$ and $F^I_{\mu \nu}=
\partial_\mu A^I_\nu - \partial_\nu A^I_\mu$ are
abelian field strength. It is easy to see that (\ref{5dAction})
reduces to $(-1)$ times the action (\ref{paraH-real}) upon setting
\[
\partial_0 \sigma^I = 0 \;,\;\;
\partial_0 A_m^I =0 \;,\;\;\;
F^I_{mn}=0 \;,
\]
identifying $b^I = A^I_0$ and dropping the integration over time. 
This type of reduction corresponds to the restriction to static and purely
electric five-dimensional backgrounds. 
As indicated by the `dots' in (\ref{5dAction}),
the five-dimensional theory could have further terms as long as 
they do not contribute to static, purely electric field configurations
involving the scalars and gauge fields. 
For example, the action for five-dimensional vector multiplets
\cite{EucI}
contains a Chern-Simons term and fermionic terms, but these
do not contribute to backgrounds which are static and where
only scalars and electric field strength are excited. Note that there is
a conventional minus sign between (\ref{5dAction}) and 
(\ref{paraH-real}). Our conventions 
for Lorentzian actions are that the space-time metric is of
the `mostly plus' type, and that kinetic terms are positive definite. 
The convention for Euclidean actions is that the 
terms for the scalars $\sigma^I$ are positive definite, while
scalar fields obtained by temporal reduction of gauge fields
have a negative definite action.

We can also reduce the five-dimensional action (\ref{5dAction})
over a space-like direction, leading to the following sigma
model on four-dimensional Minkowski space:
\begin{equation}
\label{4dactionMink}
S[\sigma, b]_{(1,3)} 
=  - \int d^4 x \;\frac{1}{2} N_{IJ}(\sigma) \left( \partial_{\bar{m}} \sigma^I
\partial^{\bar{m}} \sigma^J + \partial_{\bar{m}} b^I \partial^{\bar{m}} b^J \right) \;,
\end{equation}
where $\bar{m}=0,1,2, 3$.
Note that we have discarded all terms in (\ref{5dAction})
which do not contribute to the scalar sigma model. By a Wick 
rotation we obtain the Euclidean action
\begin{equation}
\label{Herm-real}
S[\sigma,b]'_{(0,4)} 
= \int d^4 x \frac{1}{2} N_{IJ}(\sigma) \left( \partial_m \sigma^I
\partial^m \sigma^J + \partial_m b^I \partial^m b^J \right) \;,
\end{equation}
which is positive definite.
Comparison to (\ref{paraH-real})
shows explicitly that dimensional 
reduction over space followed by a Wick rotation is different
from dimensional reduction over time. However, the two actions
(\ref{paraH-real}) and (\ref{Herm-real}) are related by the
analytic continuation $b^I \rightarrow i b^I$. In other words,
two actions obtained by space-like and by time-like reduction respectively are
related by a modified Wick rotation which acts non-trivially on
axionic scalars.

For later reference, let us introduce the following notation for
the target spaces of the actions we have encountered so far.
The five-dimensional action (\ref{5dAction}) has an $n$-dimensional
target space $M_r$ with postive definite metric $N_{IJ}$.
The four-dimensional actions (\ref{4dactionMink}) and (\ref{Herm-real})
have a $2n$-dimensional target space $M'$ with positive definite metric
$N_{IJ} \oplus N_{IJ}$, while (\ref{paraH-real}) has a $2n$-dimensional
target space $M$ with split signature metric $N_{IJ} \oplus (-N_{IJ})$.

The manifolds $M$ and $M'$ carry additional structures.
For $M'$ 
we can define complex coordinates
\[
Y^I = \sigma^I + i b^I  \;,
\]
and we see that the target space $M'$ is Hermitean:
\begin{equation}
\label{SY}
S[Y]'_{(0,4)} = \int d^4 x \; \frac{1}{2} N_{IJ}(Y+\overline{Y}) \partial_m Y^I 
\partial^m \overline{Y}^J  \;.
\end{equation}
This begs the question whether there is a similar additional 
structure for the indefinite target space $M$. And indeed, here one
can define para-complex coordinates by
\[
X^I = \sigma^I + e b^I \;,
\]
where the para-complex unit has the properties
\[
e^2 = 1 \;,\;\;\;\overline{e} = -e \;.
\]
The theory of para-complex manifolds 
runs to a large extent parallel to the theory of complex
manifolds. In particular, the concepts of para-Hermitean, 
para-K\"ahler, and special para-K\"ahler manifolds are 
analogous to their complex counterparts. We refer to
\cite{EucI,EucII,EucIII} for a detailed account. Using para-complex
coordinates, one sees that the action (\ref{paraH-real})
has a  para-Hermitean target space:
\begin{equation}
\label{SX}
S[X]_{(0,4)} = \int d^4 x \; \frac{1}{2} N_{IJ}(X+\overline{X}) \partial_m X^I 
\partial^m \overline{X}^J  \;.
\end{equation}
Thus actions of the type (\ref{paraH-real}) have a target 
space which is para-Hermitean and has 
$n$ commuting isometries acting as shifts. The latter
implies that $M$ can be obtained from an $n$-dimensional 
manifold $M_r$ with positive definite metric, by applying temporal
dimensional reduction to the corresponding action. 

The two real actions (\ref{SY}) and (\ref{SX}) can be viewed as two
different real forms of one underlying complex action. 
This is further explained in Appendix A. Complex actions 
are useful to get a more unified picture actions and solutions
which are related by analytic continuation. In \cite{Bergshoeff:Complex}
complex actions for the ten-dimensional and eleven-dimensional
maximal supergravity theories have been used to give a unified
description of domain wall and cosmological solutions. There seems
to be a close relation to the concept of fake supersymmetry. 
Complex actions seem also to be useful in understanding the
Euclidean action of four-dimensional supergravity theories \cite{EucIII}.

\subsection{From harmonic maps to harmonic functions}

Solving the equations of motion for a sigma 
model is equivalent to constructing a harmonic map 
from the space-time $X$ to the scalar target space $M$. In general,
both $X$ and $M$ can be (pseudo-)Riemannian manifolds.
We restrict
ourselves to the case where $X$ is Euclidean space $E$ equipped
with its standard flat metric. Then the action of a general sigma
model takes the form
\[
S[\Phi]_{(0,4)} = 
\int d^4 x N_{ij}(\Phi) \partial_m \Phi^i \partial^m \Phi^j \;,
\]
and the equations of motion can be brought to the form
\begin{equation}
\label{HarmonicMap}
\Delta \Phi^i + \Gamma_{jk}^i \partial_m \Phi^j \partial^m \Phi^k =0 \;,
\end{equation}
where $\Gamma^i_{jk}$ are the Christoffel symbols of the metric 
$N_{ij}$ of $M$.
This is the coordinate form of the equation of a harmonic map 
$\Phi : E \rightarrow M$ from Euclidean space $E$ to the 
(pseudo-)Riemannian target $M$.

One strategy for constructing such maps\footnote{See 
\cite{Stelle,Bergshoeff:Geodesic,EucIII} for
a more detailed review.} is to identify totally
geodesic submanifolds $N \subset M$. A submanifold $N\subset M$
is called completely geodesic if every geodesics of $N$ is also 
a geodesic of $M$. Then the embedding of $N$ into $M$ is 
a totally geodesic map, and 
since the composition of a 
harmonic map $E \rightarrow N$ 
with a totally geodesic map $N \rightarrow M$ is harmonic, it 
suffices to find harmonic maps $\phi: E \rightarrow
N \subset M$ in order to solve the scalar equations of motion.
We are interested in a criterion which guarantees 
that the solution of the 
harmonic map equation (\ref{HarmonicMap}) can be expressed in terms of
harmonic functions. This will happen in particular 
if the submanifold $N$ is flat, so that the 
Christoffel symbols vanish identically if we use affine coordinates.
Then we can parametrize the scalar fields such that the
independent scalars $\phi^a$, $a=1, \ldots, \dim N$ 
corresponds to affine coordinates on $N$, and 
the harmonic map equation reduces to
\begin{equation}
\Delta \phi^a =0 \;.
\end{equation}
If $N$ has $\dim N < \dim M$, then the solution for the 
remaining $\dim M - \dim N$ scalar fields can be expressed in terms
of the solution for the $\phi^a$. The dimension of $N$ controlls
the number of independent harmonic functions which occur in the
solution.

We will now investigate under which conditions
the reduction of the equations of motion to decoupled 
harmonic equations can be achieved, assuming that the target manifold
$M$ is para-Hermitean and has $n$ commuting shift symmetries. 
In this case it is convenient to write the equations of motion in terms
of the real fields $\sigma^I$, $b^I$. By variation of the
action (\ref{paraH-real}) we obtain:
\begin{eqnarray}
\partial^m \left( N_{IJ} \partial_m \sigma^J \right) - \frac{1}{2}
\partial_I N_{JK} \left( \partial_m \sigma^J \partial^m \sigma^K 
- \partial_m b^J \partial^m b^K \right) &=& 0 \;,\nonumber \\
\partial^m \left( N_{IJ} \partial_m b^J \right) &=& 0\;. \label{FullEOM}
\end{eqnarray}
This could be cast into the form (\ref{HarmonicMap}),
but in the present  form it is manifest that a drastic simplification occurs
if we impose that 
\begin{equation}
\label{InstantonAnsatz}
\partial_m \sigma^I = \pm \partial_m b^I \;.
\end{equation}
In this case the two equations (\ref{FullEOM}) collapse into 
\begin{equation}
\label{ReducedEOM}
\partial^m \left( N_{IJ} \partial_m \sigma^J \right) = 0 \;,
\end{equation}
which is very close to the harmonic equation. We will refer to the
condition (\ref{InstantonAnsatz}) as the extremal 
instanton ansatz. Geometrically,
the extremal instanton ansatz implies that the scalar fields are restricted 
to vary along the null directions of the metric of $M$. In other words,
the scalars take values in a submanifold $N\subset M$ which is 
completely isotropic. 

The extremal instanton ansatz has the consequence 
that the energy momentum tensor vanishes 
identically. The `improved', symmetric energy momentum tensor for the
action (\ref{paraH-real}) is obtained by variation with respect to 
a Riemannian background metric on $E$:
\begin{equation}
\label{EMtensor}
T_{mn} = N_{IJ} \left( \partial_m \sigma^I \partial_n \sigma^J -
\partial_m b^I \partial_n b^J \right) - \frac{1}{2} \delta_{mn}
N_{IJ} \left( \partial_l \sigma^I \partial^l \sigma^J - 
\partial_l b^I \partial^l b^J \right) \;.
\end{equation}
Since we are in four dimensions (more generally in $>2$ dimensions),
the vanishing of $T_{mn}$ is equivalent to
\begin{equation}
\label{NullDirections}
N_{IJ} \left( \partial_m \sigma^I \partial_n \sigma^J -
\partial_m b^I \partial_n b^J \right) =0 \;,
\end{equation}
which means that the scalar fields vary along the null directions of
$N_{IJ} \oplus (-N_{IJ})$. More precisely, depending on the choice
of sign in (\ref{InstantonAnsatz}), the scalar fields vary along
the eigendirection of the para-complex structure with eigenvalue
$(+1)$ or $(-1)$, respectively. Thus the submanifold
$N$ is the integral manifold of an eigendistribution of the
para-complex structure.

The vanishing of the energy momentum tensor has the important
consequence that solutions of (\ref{paraH-real}) which satisfy
(\ref{InstantonAnsatz}) remain solutions, without any modification,
if we couple the sigma model to gravity,
\begin{equation}
\label{SigmaR}
S[g,\sigma,b]_{(0,4)} = \int d^4 x \sqrt{g} \;
\frac{1}{2} \left(  -R  + N_{IJ} \partial_m \sigma^I \partial^m \sigma^J
- N_{IJ} \partial_m b^I \partial^m b^J \right) \;.
\end{equation}

Since the energy momentum
tensor vanishes, it is consistent to solve the Einstein equation
by taking the metric to be flat, $g_{mn} = \delta_{mn}$. 
Thus the instanton solutions we find 
are solutions of sigma models coupled to gravity (\ref{SigmaR}), subject to 
the `Hamiltonian constraint' $T_{mn}=0$. As 
we will discuss in more detail later, instanton solutions of
(\ref{paraH-real}) can therefore be lifted consistently to 
solutions of five-dimensional gravity coupled to matter. In this
case the lifting works somewhat differently than in the rigid case,
because one has to take into account the Kaluza-Klein scalar.
The resulting five-dimensional solutions are not flat, but 
have a conformally flat four-dimensional part, as is typical for
BPS solutions. One particular class of lifted solutions are 
extremal static five-dimensional black holes. The theories and
solutions obtainable from (\ref{paraH-real}) include 
all five-dimensional supergravity theories with abelian vector
multiplets and their BPS black hole solutions.
We will refer to instanton solutions obtained by the extremal instanton
ansatz as extremal. One reason for this choice of terminology 
is that they can be lifted to
extremal black hole solutions. Another reason is the saturation 
of a Bogomol'nyi bound for the action, which will be discussed in 
Section 3.

The indefiniteness of the metric of $M$ is essential 
for obtaining non-trivial solutions of the scalar
equations of motion with vanishing energy momentum
tensor. For a positive (or negative) definite scalar target space metric,
$T_{mn}=0$ would imply that all scalar fields have to
be constant.

The extremal 
instanton ansatz (\ref{InstantonAnsatz}) is sufficient but not necessary
for the vanishing of the energy momentum tensor and the reduction
of the equations of motion to (\ref{ReducedEOM}).
If the metric $N_{IJ}$ is invariant under transformations of the 
form
\begin{equation}
\label{RotIsometry}
N_{IJ} \rightarrow N_{KL} R^K_{\;\;I} R^L_{\;\;J}  \;,
\end{equation}
where $R^I_{\;\;J}$ is a constant matrix, already the generalized 
instanton ansatz\footnote{`Generalized extremal instanton ansatz' would
be more accurate, but we will use `generalized instanton ansatz' for
convenience.} (\ref{InstantonAnsatz}) 
\begin{equation}
\label{GenInstantonAnsatz}
\sigma^I = R^I_{\;\;J} b^J \;
\end{equation}
implies $T_{mn}=0$ and (\ref{ReducedEOM}).
Geometrically, the transformation (\ref{RotIsometry})
corresponds to an isometry 
of $N_{IJ} \oplus (-N_{IJ})$ which acts by
\[
\sigma^I \rightarrow \sigma^I \;,\;\;\;
b^I \rightarrow R^I_{\;\;J} b^J \;.
\]
The relation (\ref{GenInstantonAnsatz}) has occured previously 
in the context of extremal black hole solutions
of supergravity, where $R^I_{\;\;J} \not= \delta^I_J$ corresponds
to non-BPS solutions \cite{CerDal,CCDOP}.
The simplest examples of non-BPS black holes correspond
to flipping some of the charges of the black hole, which 
corresponds to diagonal $R$-matrices with entries $\pm 1$. 
Geometrically, this means that some of the fields vary along
the $(+1)$-eigendirections of the para-complex strucure while
the rest varies along the $(-1)$-eigendirections. In this
way the distinction between BPS and non-BPS extremal solutions
in supergravity can be understood geometrically and 
extended to a larger class of non-supersymmetric theories.

Let us now investigate the reduced equations of motion
(\ref{ReducedEOM}), which remain to be solved after imposing
the extremal instanton ansatz (\ref{InstantonAnsatz}) or its
generalization (\ref{GenInstantonAnsatz}):
\[
\partial^m \left( N_{IJ} \partial_m \sigma^J \right) = 0 \;.
\]
This reduces to a set of $n$ harmonic equations, provided 
there exist `dual fields' $\sigma_I$ with the property
\[
\partial_m \sigma_I = N_{IJ} \partial_m \sigma^J \;.
\]
The existence of such dual fields implies the integrability
condition
\[
\partial_{[n} ( N_{IJ} \partial_{m]} \sigma^J) = 0  \;.
\]
The same condition has been observed in the
context of five-dimensional black hole solutions in \cite{PSVV:08}.
Since
$\partial_{[n} \partial_{m]} \sigma^J = 0$, the integrability condition is
equivalent to
\begin{equation}
\label{ConditionOnN}
\partial_{[n}  N_{IJ} \partial_{m]} \sigma^J = 
\partial_K N_{IJ} \partial_{[n} \sigma^K \partial_{m]} \sigma^J =
0 \;.
\end{equation}
There are two strategies for solving this constraint.
The first is to restrict the solution $\sigma^I(x)$ while not 
making assumptions about the metric $N_{IJ}$. If we assume
that the solution only depends on one of the coordinates
of $E$, then (\ref{ConditionOnN}) is solved automatically. The most natural
assumption is spherical symmetry, $\sigma^I = \sigma^I(r)$, 
where $r$ is a radial coordinate, as this admits solutions
which asymptotically approach ground states $\sigma^I_{\rm vac} = \mbox{const}$
at infinity. In this case the explicit solutions of $\Delta \sigma_I=0$
are single-centered harmonic functions,
\[
\sigma_I = H_I(r) = h_I + \frac{q_I}{r^2}  \;,
\]
where $h_I$ and $q_I$ are constants. The constants $h_I$ 
specify the values of $\sigma_I$ at infinity. As we will see
in the next section the parameters $q_I$ are charges. Such solutions
can be interpreted as describing an instanton with charges
$q_I$ located at $r=0$.\footnote{At $r=0$ the fields $\sigma_I$ 
take singular values, and the equations of motion (\ref{ReducedEOM})
are not satisfied, unless explicit source terms are added. This
is analogous to electric point charges in electrostatics.} Geometrically,
this type of solution corresponds to a situation where the
scalars flow along a null
geodesic curve in $M$.

The second strategy is to make no assumption about the 
solution. This is compulsory if we want that there are multi-centered
solutions, 
\begin{equation}
\label{SolDualsigma}
\sigma_I(x) = H_I(x) = h_I + \sum_{a=1}^N \frac{q_{aI}}{|x - x_a|^2}\;,
\end{equation}
where $h_I, q_{aI}$ are constants and where $x,x_a\in E$.
Such solutions correspond to $N$ instantons with 
charges $q_{aI}$, which are located at the positions $x_a$. 
For multi-centered solutions we cannot impose spherical 
symmetry but the integrability condition (\ref{ConditionOnN})
can still be solved
by imposing the condition
\begin{equation}
\label{HesseMetric1}
\partial_{[K} N_{I]J} = 0  
\end{equation}
on the scalar metric. This is equivalent to requiring that
the first derivatives $\partial_K N_{IJ}$ of the metric
are completely symmetric, or, again equivalently, that the
Christoffel symbols of the first kind $\Gamma_{IJ|K}$ are
completely symmetric. Finally, by applying the Poincar\'e 
lemma twice, we see that (\ref{HesseMetric1}) is locally
equivalent to the existence of a Hesse potential ${\cal V}(\sigma)$:
\begin{equation}
\label{HesseMetric}
N_{IJ} = \frac{\partial^2 {\cal V}}{\partial \sigma^I \partial \sigma^J} \;.
\end{equation}
A coordinate-free formulation is obtained by observing that 
the local existence of a Hesse potential is equivalent to the
existence of a flat, torsion-free connection $\nabla$ which
has the property that $\nabla g$, where $g$ is the metric, is
a completely symmetric rank 3 tensor field. This is the definition 
of a Hessian metric given in \cite{AleCor}. They also observed 
that the affine special real manifolds which are the target spaces
of rigid five-dimensional vector multiplets are special 
Hessian manifolds where the cubic form $\nabla g$
is parallel with respect to $\nabla$. It is easy to see
why supersymmetry requires this additional condition. 
Supersymmetry implies the presence of a Chern-Simons term,
whose coefficient is given by $\nabla g$. 
Gauge invariance requires that this coefficient is covariantly 
constant. In affine coordinates, this becomes the well
known condition that the third derivatives of the Hesse potential
(which for rigid supersymmetry is identical with the prepotential)
must be constant. Hence the Hesse potential must be a cubic 
polynomial. In this paper we consider more general Hesse 
potentials, but since the models are not supersymmetric, 
there is no fixed relation between the Chern Simons term (if any is
present) and other terms in the Lagrangian. In the purely 
electric background that we consider, a Chern-Simons does not
contribute, and therefore we do not need to investigate whether
a Chern-Simons term could or should be added.

The dimensional reduction of models with general Hessian
target spaces leads to a generalization 
of the rigid version of the $r$-map \cite{AleCor}. Recall that the $r$-map 
relates the target spaces of five-dimensional and four-dimensional
vector multiplets \cite{dWvP:1992}. 
The $r$-map can be derived 
by dimensionally reducing the vector multiplet action from five
to four dimensions, and depending on whether one considers 
supersymmtric field theories or a supergravity theories one 
obtains a rigid (also called affine) or local (also called projective) 
version of
the $r$-map. Affine (projective) very special real manifolds
are mapped to affine (projetive) special K\"ahler manifolds,
respectively. The generalized rigid $r$-map of \cite{AleCor} is
obtained by relaxing the constraint that the scalar target geometry
of the five dimensional theory is very special real and only 
requiring it to be Hessian. In the notation of our paper the
resulting generalized $r$-map is obtained by reducing (\ref{5dAction})
to (\ref{Herm-real}) while imposing that $N_{IJ}(\sigma)$ 
satisfies (\ref{HesseMetric}). As shown in \cite{AleCor}
the resulting target space $M'$ of the four-dimensional theory
is a K\"ahler manifold with $n$ commuting shift isometries. 
We already noted that $M'$ is Hermitean. To check that it is
K\"ahler we go to complex coordinates $Y^I = \sigma^I + i b^I$ 
and verify by explicit calculation that 
\[
K(Y,\bar{Y}) = K(Y+\bar{Y}) = 4 {\cal V}(\sigma(Y,\bar{Y})) \;
\]
is a K\"ahler potential for the metric $N_{IJ} \oplus N_{IJ}$
of $M'$. Note that \cite{AleCor} prove that the relation 
between Hessian manifolds $(M_r, N_{IJ})$ and K\"ahler manifolds
$(M',N_{IJ} \oplus N_{IJ})$ is one-to-one: any K\"ahler 
manifold with $n$ commuting shift isometries can be obtained 
from a Hessian manifold by the generalized $r$-map. 

By reducing (\ref{5dAction}), with Hessian $N_{IJ}$, over 
time rather than space, we obtain (\ref{paraH-real})
and a 
para-version (or temporal
version) of the generalized rigid $r$-map. As shown in 
\cite{EucI}, the rigid para-$r$-map relates affine very special real 
manifolds to affine special para-K\"ahler manifolds with a 
cubic prepotential. If we only impose that $M_r$ is Hessian,
then $M$ is a para-K\"ahler manifold with $n$ commuting 
shift isometries. We have already seen that metric $N_{IJ} 
\oplus (- N_{IJ})$ is para-Hermitean. To see that it is
para-K\"ahler we go to para-complex coordinates $X^I=\sigma^I
+ e b^I$ and verify that 
\[
K(X,\bar{X}) = K(X+\bar{X}) = 4 {\cal V}(\sigma(X,\bar{X})) \;
\]
is a para-K\"ahler potential. We expect that
every para-K\"ahler metric with $n$ commuting shift isometries
can be obtained in this way.

To conclude this section, let us discuss how our class of
solutions fits into the general set up 
of constructing harmonic maps $E\rightarrow M$ by finding
totally geodesic embeddings $N \subset M$ and harmonic 
maps $E \rightarrow N$. The extremal instanton ansatz implies that
in our construction $N$ is totally isotropic.
Although the induced metric of $N$ is totally degenerate
one can still construct harmonic maps $E \rightarrow M$,
by decomposing `harmonic maps'\footnote{Strictly speaking, 
we should not call this a harmonic map if $N$ is totally isotropic,
because the definition of a 
harmonic map requires that source and target manifolds are equipped
with non-degenerate metrics. However, the relevant point is that
the composed map $E \rightarrow N \subset M$ is harmonic.}
 $E\rightarrow N$ with 
totally geodesic embeddings $N \subset M$ \cite{EucIII}.
Our explicit calculation using the coordinates $\sigma^I, b^I$ 
shows that the totally isotropic submanifolds defined by the 
extremal instanton ansatz must be totally geodesic. This can also be 
understood as follows.  
The submanifolds defined by the ansatz (\ref{InstantonAnsatz})
are eigendistributions of the
para-complex structure of $M$. The integrability condition 
(\ref{HesseMetric1}) implies that $M$ is 
para-K\"ahler, and therefore the eigendistributions 
are integrable and parallel with respect to the Levi-Civita
connection. Such submanifolds are in particlar totally geodesic.

By explicit calculation we have also seen that $N$ must 
be flat. This can again be understood geometrically. 
In complete analogy to K\"ahler manifolds, the
Riemann tensor of a para-K\"ahler manifold has a particular 
index structure, when written in para-complex coordinates.
The only non-vanishing components are those where both 
pairs of indices are of mixed type, i.e. one para-holomorphic
and one anti-para-holomorphic index. However, the pullback of
the Riemann tensor to an eigendistribution of the para-complex
structur is of pure type, i.e. the non-vanishing components
have either only  
para-holomorphic or only anti-para-holomorphic indices. Therefore
the pull back of the Riemann tensor onto these eigendistributions
vanishes. Thus the pulled back connection is flat, and the
harmonic map equations must reduce to harmonic equations when
expressed in affine coordinates. We can also understand why
the existence of single-centered solutions does not impose 
constraints on the scalar target metric. In this case $N$ is
a null geodesic curve, and therefore it is flat 
for any choice of the metric of $M$.

The additional feature which is required for the existence 
of multi-centered solutions is the existence of a potential.
For many purposes it is convenient to use real coordinates
$\sigma^I, b^I$ and to work with the Hesse potential ${\cal V}(\sigma)$.
The affine coordinates on $N$, in terms of which the 
equations of motion reduce to decoupled harmonic equations
$\Delta \sigma_I=0$ are given by the first derivatives of the
Hesse potential 
\begin{equation}
\label{DualCoordinate}
\sigma_I \simeq \frac{\partial {\cal V}}{\partial \sigma^I} \;.
\end{equation}
This is clear 
because the application of the partial derivative $\partial_m$ 
on (\ref{DualCoordinate})
gives the integrability
condition (\ref{ConditionOnN}). In (\ref{DualCoordinate})
we have left the constant proportionality undetermined, so that
we can later fix it to convenient numerical values case by case.
Given the Hesse potential we have an explicit formula for
the dual fields $\sigma_I$ in terms of the scalars $\sigma^I$. 
In general it is not possible to give explicit expressions
for the scalars $\sigma^I$ as functions of the dual scalars
$\sigma_I$. Hence, while instanton solutions are completely 
determined by harmonic functions $H_I$ through a set of
algebraic relations, it is not
possible in general to express the solution 
in terms of the $H_I$ in closed form.
However, we will see that this does not
prevent us from understanding many features of the solutions.
Moreover, if the Hesse potential is sufficiently simple, 
explicit expressions can be obtained. 
Examples will be given later.

\subsection{From instanton charges to harmonic functions}

We have already anticipated that the parameters $q_I$ or
$q_{Ia}$ occuring in the harmonic functions $H_I$ can be
interpreted as charges. In this section we provide the
definition of instanton charges and derive the instanton
solutions from a slightly different perspective. It has been
observed in the literature on extremal non-BPS black holes
that if solutions can be expressed in terms of harmonic functions
then the equations of motion can often be reduced from second
to first order equations \cite{CerDal,CCDOP,PSVV:08}. 
Our derivation will show that these
two properties are related through the existence of conserved
charges: imposing that solutions carry {\em finite} instanton charges
implies that the equations of motion can be replaced by 
first order equations. 

The symmetry of the target manifold $M$ under constant
shifts $b^I \rightarrow b^I + C^I$ implies the existence
of $n$ charges, which we call instanton charges. As we will
see later these lift to five-dimensional electric charges.
The current associated to the shift symmetry is 
\[
j_I = \partial_m \left(  N_{IJ}(\sigma)  \partial^m b^J \right) \;.
\]
It is `conserved' in the sense that the Hodge-dual four-form 
is closed. The charge obtained by integrating 
this current over four-dimensional Euclidean space is 
\begin{equation}
\label{DefInstCharge}
Q_I = \int d^4x j_I \;.
\end{equation}
Since $j_I$ is a total derivative, the charge $Q_I$
can be re-written as a surface charge, as is typical for gauge theories:
\[
Q_I = 
\oint d^3 \Sigma^m \left(  N_{IJ}(\sigma)  \partial_m b^J \right) \;.
\]
The integral is performed over a topological three-sphere which 
encloses all sources. Note that 
explicit sources are needed to have
non-vanishing instanton charges, because 
the equation of motion (\ref{FullEOM})
for $b^I$ implies $j_I =0$. To obtain non-trivial solutions we
allow the presence of pointlike ($\delta$-function type)
sources. 
Pointlike sources in 
Euclidean theories are referred to as $(-1)$-branes. The existence
of solutions carrying $(-1)$-brane charge is taken as 
evidence that the theory should be extended by adding $(-1)$-branes.
This is the philosophy underlying field theory and 
string dualities, for which 
our class of actions provides models.

Solutions with non-vanishing instanton charge must have a particular
asymptotic behaviour. We assume that the sources are contained in a finite
region and take the limit $r\rightarrow \infty$, where $r$ is
a radial coordinate with origin within this region. 
The integrand can be expanded in powers of $\frac{1}{r}$, and
we assume 
that the contribution of the leading term to the charges 
$Q_I$ is non-vanishing and finite.\footnote{The expansion in $\frac{1}{r}$ 
is of course
a version of the multipole expansion. In fact, from the five-dimensional
point of view it is literally the multipole expansion of a discrete charge 
density contained in a finite region.}  
This implies that subleading 
terms in $\frac{1}{r}$ do not contribute to the charge.
Since the leading term is independent of the angles, we take
the integration surface
to be a three-sphere $S^3_r$ of radius $r$ and
integrate over the angles. The resulting charge is
\begin{equation}
\label{InstChargeMulti}
Q_I =  \mbox{vol}(S^3_1) \lim_{r \rightarrow \infty} \left(
r^3 N_{IJ}(\sigma) \partial_r b^J \right) \;,
\end{equation}
where $\mbox{vol}(S^3_1) = 2 \pi^2$ is the 
the volume of the unit three-sphere. 
Since we assume that $Q_I$ is neither infinite nor zero, it follows
that the integrand 
$N_{IJ} \partial_r b^J$ falls off like $\frac{1}{r^3}$:
\[
N_{IJ} \partial_r b^J =  \frac{1}{2\pi^2} \frac{Q_I}{r^3} + \cdots \;,
\]
where 
the omitted terms are of order $\frac{1}{r^4}$.
Now we observe that the leading term
in $N_{IJ} \partial_r b^J$ is the derivative of a spherically symmetric
harmonic function $\tilde{H}_I(r)$:
\begin{equation}
\label{AsymptoticSol}
N_{IJ} \partial_r b^J = \frac{1}{2\pi^2} \frac{Q_I}{r^3} + \cdots 
= \partial_r \tilde{H}_I(r) + \cdots \;,
\end{equation}
with
\[
\tilde{H}_I(r) = \frac{\tilde{q}_I}{r^2} + \tilde{h}_I \;,
\]
where $\tilde{h}_I$ are constants and where $\tilde{q}_I =
-\frac{1}{4\pi^2} Q_I$ are proportional to the charges.
For simplicity we will refer to the parameters $\tilde{q}_I$
as charges.

While the leading term of the expanded solution is automatically
spherically symmetric, the full solution may not be. 
This leads to the distinction of two cases,
precisely as in the previous section. 
If we impose that the full solution is spherically symmetric, then
we obtain a solution to the equations of motion by setting 
all subleading terms to zero, and imposing the extremal instanton ansatz
(\ref{InstantonAnsatz}) or its generalized version (\ref{GenInstantonAnsatz}):
\[
N_{IJ} \partial_r b^J = \partial_r \tilde{H}_I(r)  \;.
\]
This solution is spherically symmetric and at $r=0$ the equations
of motion need to be modified by a $\delta$-function type source term
with coefficient $q_I$. The source is interpreted as a 
$(-1)$-brane of total charge $\tilde{q}_I$, which is located at the center
$r=0$ 
of the harmonic function. 

If we do not assume  that the full solution is spherically  symmetric, 
then we need to find solutions of (\ref{ReducedEOM}) with asymptotics 
(\ref{AsymptoticSol}) subject to the (generalized) extremal instanton ansatz.
Such solutions are obtained if
\begin{equation}
\label{SolDualb}
N_{IJ} \partial_m b^J = \partial_m \tilde{H}_I(x)\;,
\end{equation}
where $\tilde{H}_I(x)$ are harmonic functions. Since the
right hand side is a total derivative, we need to impose
an integrability condition equivalent to (\ref{HesseMetric1}),
and thus we recover
the condition that the scalar metric $N_{IJ} \oplus (-N_{IJ})$ 
of $M$ must be a para-K\"ahler metric.  Assuming this, we have managed
to reduce the second order equations of motion (\ref{ReducedEOM})
to the first order quasilinear partial differential equations
\begin{equation}
\label{1stOrderForm}
\partial_m b^I = N^{IJ} \partial_m \tilde{H}_J(x) \;,
\end{equation}
where $N^{IJ}$ is the inverse of $N_{IJ}$. From our derivation
it is clear that the crucial ingredient for the reduction of the
order of the equation of motion is
the existence of the charges $Q_I$, which can be used to
prescribe the asymptotic behaviour of the solution, and 
`to peel off' one derivative from the equation of motion,
provided that the integrability condition (\ref{ConditionOnN})
holds.

Solutions with the correct asymptotics are given by multi-centered
harmonic functions
\begin{equation}
\tilde{H}_I(x) = \tilde{h}_I + \sum_{a=1}^N  \frac{\tilde{q}_{aI}}{|x - x_a|^2}
\end{equation}
where $x,x_a \in \mathbbm{R}^4$. They correspond to $N$ $(-1)$-branes
with charges $\tilde{q}_{aI}$, which located at the centers $x_a$. For 
$|x| \rightarrow \infty$ the leading term is 
\begin{equation}
\tilde{H}_I (x) \approx  \frac{1}{|x|^2} \sum_{a=1}^N \tilde{q}_{aI} 
+ {\cal O}(|x|^{-3}) \;.
\end{equation}
Thus the total instanton charges of such a configuration are 
$\tilde{q}_I = \sum_{a=1}^N \tilde{q}_{aI}$.

The relation between this version of the solution and the one
given in the previous section 
is provided by the (generalized) extremal instanton ansatz.
Observe that 
\[
\partial_r \sigma_I = N_{IJ} \partial_r \sigma^J = N_{IJ} R^J_{\;\;K}
\partial_r b^K = R_I^{\;\;J} N_{JK} \partial_r b^K \;,
\]
where $R_I^{\;\;J}$ is the inverse of the transposed of $R^I_{\;\;J}$.
Comparing the solutions (\ref{SolDualsigma}) and (\ref{SolDualb})
we conclude
\[
\partial_m H_I = R_I^{\;\;J} \partial_m \tilde{H}_J \;,
\]
which implies $H_I = R_I^{\;\;J} \tilde{H}_J$ up to 
an additive constant. This constant reflects the 
shift symmetry $b^I \rightarrow b^I + C^I$. 
However the coefficients of the non-constant terms 
in the harmonic functions are not ambiguous and related
by the rotation matrix $R_I^{\;\;J}$. In particular the
instanton charges $q_I$ and $\tilde{q}_I$ are related by 
\[
q_I = R_I^{\;\;J} \tilde{q}_J \;.
\]
Thus we have seen that the reduction of the equations of motion to
the decoupled harmonic equations and the reduction of the 
equations of motion from second to first order differential equations
result from the same
integrability condition (\ref{ConditionOnN}).
The integrability condition can be solved by either imposing
that the solution is spherically symmetric, or by restricting
the target geometry to be para-K\"ahler.

\subsection{Spherically symmetric solutions and flow equations}

In this section we take a closer look at spherically symmetric
solutions. 
For spherically
symmetric black hole solutions (and other related solitonic
solutions), the reduction of the equation of motion to 
first order equations was first noticed for
BPS solutions. In this context the first order equations are
known as (generalized) attractor equations, or (gradient) flow 
equations.
Later it was realized that a first order rewriting 
is often also possible for non-BPS solutions, and leads to first
order flow equations which are driven by potential, which generalizes
the $N=2$ central charge \cite{CerDal,CCDOP,PSVV:08}.
Let us therefore explain how
gradient flow equations fit into our framework.

We have seen previously that if we impose that solutions carry 
non-vanishing instanton charge (and satisfy the integrability 
condition (\ref{ConditionOnN}) which is trivial for spherically
symmetric solutions), then the second order equations of motion 
reduce to the first order quasilinear partial differential equations
(\ref{1stOrderForm}).   If we impose spherical symmetry 
the equations  reduce further to the first order quasilinear
ordinary differential equations
\[
\sigma^I{}' = N^{IJ}(\sigma)  H'_J(r) = 
N^{IJ}(\sigma)  \frac{d}{dr} \left( \frac{q_J}{r^2} + h_J \right) \;,
\]
where $f'=\frac{df}{dr}$.
The standard form of the flow equations is obtained by introducing
the new coordinate $\tau=\frac{1}{r^2}$:
\[
\dot{\sigma}^I = N^{IJ}(\sigma)  \frac{d}{d\tau}
\left( q_J \tau + h_J \right) = N^{IJ} q_J \;,
\]
where $\dot{f}=\frac{df}{d\tau}$.
By introducing the function 
\[
W = q_J \sigma^J
\]
and obtains the gradient flow equations
\begin{equation}
\label{1storder_gradient_flow}
\dot{\sigma}^I  = N^{IJ} \partial_J W \;.
\end{equation}

In terms of the
instanton charges $Q_I \propto \tilde{q}_I$, the `superpotential' is
\[
W = R_I^{\;\;J} \tilde{q}_J \sigma^I\;.
\]
This form is familiar from black holes \cite{CerDal,CCDOP,PSVV:08}.
If the underlying theory is supersymmeric, and if the 
$R$-matrix is proportional to the identity, then 
\[
W = \pm Z = \pm \tilde{q}_I \sigma^I \;,
\]
where $Z$ is the real central charge of the supersymmetry algebra
of the underlying five-dimensional theory. $Z$ is
also one of the two real supercharges of the four-dimensional 
Euclidean supersymmetric theory obtained by reduction over time.

The new coordinate $\tau$ has a simple geometrical interpretation.
To see this consider the version of the equations of motion which
involve the dual scalars $\sigma_I$:
\[
\Delta \sigma_I =0 \;.
\]
This is the harmonic equation for a map from space-time $E$ into a flat
submanifold $N\subset M$, written in terms of affine coordinates 
on $N$. For spherically symmetric solutions, this takes the form
\[
\left( \frac{\partial^2 }{\partial r^2} + \frac{3}{r} 
\frac{\partial}{\partial r} \right) \sigma_I = 0 \;.
\]
This is the geodesic equation for a curve on a flat submanifold
$N\subset M$. The presence of the
second term shows that $r$ is not an affine curve parameter. 
However, one can always introduce
an affine parameter $\tau$, which is unique up to affine transformations,
such that equations reduces to 
\[ 
\frac{\partial^2 }{\partial \tau^2} \sigma_I =0 \;.
\]
It is easy to see that for the case at hand the affine parameters are
\[
\tau = \frac{a}{r^2} + b \;,
\]
where $a\not=0$ and $b$ are constants. Thus $r\rightarrow \tau=\frac{1}{r^2}$
is a reparametrization of the geodesic which brings it to affine form.
The solution takes the particularly simple form of harmonic functions
in one variable,
\[
\sigma_I (\tau) = q_I \tau + \sigma_I(0) \;.
\]

For completeness, let us review an alternative derivation of the
flow equations, which uses a variant of the Bogomol'nyi trick and
which is used frequently in the literature on non-BPS extremal black holes
(see for example \cite{CerDal,CCDOP,PSVV:08}).
In spherically symmetric backgrounds $ds^2 = dr^2 + r^2 d \Omega_{(3)}^2$ 
the action (\ref{paraH-real}) 
can be reduced to the one-dimensional action
\[
S[\sigma,b]_{(0,1)} = \int r^3 dr  \; \frac{1}{2} 
N_{IJ} \left( \sigma^I{}' \sigma^J{}'
- b^I{}' b^J{}' \right) \;.
\]
Then one tries to rewrite this action as an (alternating) sum
of perfect squares plus boundary terms. The factor $r^3$ in
can be eliminated 
by going to the affine curve parameter $\tau=\frac{1}{r^2}$:
\begin{equation}
\label{1daction}
S[\sigma,b]_{(0,1)} = \frac{1}{2} \int d \tau N_{IJ} \left( \dot{\sigma}^I
\dot{\sigma}^J - \dot{b}^I
\dot{b}^J \right) \;.
\end{equation}
The Euler-Lagrange equations of this action need to be 
supplemented by the constraint
\begin{equation}
\label{HamConstr}
{\cal H} = \frac{1}{2} N_{IJ} \left(
\dot{\sigma}^I \dot{\sigma}^J - \dot{b}^I \dot{b}^J\right) =0 \;,
\end{equation}
which implies that the solution is extremal.\footnote{In general the
constraint is $H=c^2$, where $c$ is a constant, but the case $c\not=0$
corresponds to non-extremal solutions \cite{FGK}, 
which we do not consider in this
paper.} From the five-dimensional point of view, this
constraint imposes the Einstein equations, and therefore it is analogous
to the four-dimensional constraint $T_{mn}=0$. Note that
${\cal H}$ in (\ref{HamConstr}) is the Hamiltonian of the 
one-dimensional action. The canonical momenta 
\begin{equation}
\label{LargeMomenta}
p_I= \frac{\partial {\cal L}}{\partial \dot{\sigma}^I} = 
N_{IJ} \dot{\sigma}^J \;\;\;\mbox{and}\;\;\;
\tilde{p}_I= \frac{\partial {\cal L}}{\partial \dot{b}^I} = 
N_{IJ} \dot{b}^J
\end{equation}
are conserved and agree with 
the charges:  $p_I = q_I$ and $\tilde{p}_I = \tilde{q}_I$. 
Since the Lagrangian is
quadratic in the velocities and does not contain a potential, the 
Hamiltonian coincides with the Lagrangian. 

The first order form of the equations of motion can be obtained by
rewriting the Lagrangian as an (alternating) sum of squares, up to
boundary terms \cite{CerDal,CCDOP,PSVV:08}:
\begin{eqnarray}
S[\sigma,b]_{(0,1)} &=& \frac{1}{2} \int d \tau 
\big[ N_{IJ} \left( \dot{\sigma}^I - N^{IK} q_K \right) 
\left( \dot{\sigma}^J - N^{JL} q_L \right)  \label{BogoVar} \\
&& - N_{IJ} \left( \dot{b}^I - N^{IK} \tilde{q}_K \right) 
\left( \dot{b}^J - N^{JL} \tilde{q}_L \right) \big]
+ \mbox{boundary terms} \;, \nonumber 
\end{eqnarray}
where the constants $q_I$ and $\tilde{q}_I$ are related by
$q_I = R_I^{\;\;J} \tilde{q}_J$.
Since the boundary terms do not contribute to the equations of motion,
a subclass of solutions is obtained by setting both squares 
to zero. This is equivalent to the combined 
flow equations for $\sigma^I$ and
$b^I$, or to the generalized instanton ansatz 
$\dot{\sigma}^I = R^I_{\;\;J} \dot{b}^J$
together with the
flow equations for the independent scalars.

\subsection*{Reduced scalar manifold, geodesic potential, and
remarks on the Hamilton-Jacobi formalism}

So far we have worked on the scalar manifold $M$, which is
parametrized 
by the $2n$ scalars $\sigma^I$ and $b^I$. One approach 
used frequently in the literature is to eliminate the 
$b^I$ by their equations of motion, which results in an
effective potential for the $\sigma^I$ which contains the 
charges as parameters \cite{FGK}. The resulting equation for
$\sigma$ describes geodesic motion 
with a non-trivial potential on the $n$-dimensional manifold
$M_r$. 
We briefly review this approach 
in order to explain how our work relates to \cite{ADOT},
who applied the Hamilton-Jacobi formalism to 
spherically symmetric, static black holes.\footnote{While 
\cite{ADOT} also consider non-extremal black holes, we restrict
ourselves to the extremal case.} To facilitate the comparision,
it is convenient to write the Euler-Lagrange equations of
the action (\ref{1daction}) in the following form:
\begin{eqnarray}
\ddot{\sigma}^I + \Gamma^I_{JK} \dot{\sigma}^J \dot{\sigma}^K
+ \frac{1}{2} N^{IL} \partial_L N_{JK} \dot{b}^J \dot{b}^K &=& 0 \;,
\label{1dgeodesic} \\
\frac{d}{d\tau} \left(N_{IJ} \dot{b}^J \right)&=& 0 \;. \nonumber
\end{eqnarray}
Here $\Gamma^{I}_{JK}$ are the Christoffel symbols of the
metric $N_{IJ}(\sigma)$ on the manifold $M_r$.
While the combined set of equations is
the geodesic equation for the metric $N_{IJ} \oplus (- N_{IJ})$
on the manifold $M$, one can use the fact that $N_{IJ}$ is 
independent of the $b^I$ to eliminate the $b^I$ and thus 
obtain a geodesic equation with potential on $M_r$. 
The equations of motion of the $b^I$ state that the quantities
\[
\tilde{q}_I = N_{IJ} \dot{b}^J 
\]
are conserved. In fact,  the $\tilde{q}_I$ are the conserved axionic 
charges introduced previously. Using this the equations
(\ref{1dgeodesic}) reduce to
\begin{equation}
\label{1dgeodesic_with_potential}
\ddot{\sigma}^I + \Gamma^I_{JK} \dot{\sigma}^J \dot{\sigma}^K
+ \frac{1}{2} N^{IL} \partial_L N_{JK} N^{JM} N^{KN}
\tilde{q}_M \tilde{q}_N =0 \;.
\end{equation}
The constraint (\ref{HamConstr}) now takes the form
\begin{equation}
{\cal H} = \frac{1}{2} \left( N_{IJ} \dot{\sigma}^I \dot{\sigma}^J
- N^{IJ} \tilde{q}_I \tilde{q}_J \right) =0 \;.
\end{equation}
Expressing this in terms of the canonical momenta $p_I = N_{IJ} \dot{\sigma}^J$
and defining the `geodesic potential'
\begin{equation}
V(\sigma)_{\tilde{q}} = N^{IJ} \tilde{q}_I \tilde{q}_J \;,
\end{equation}
the Hamiltonian constraint becomes
\[
\tilde{\cal H}(\sigma, p) = \frac{1}{2} \left( p_I N^{IJ} p_J 
- V(\sigma)_{\tilde{q}} \right) = 0 \;.
\]
The geodesic  potential is positive definite for positive definite
$N_{IJ}$ \cite{FGK,ADOT}. The relative minus sign between the `kinetic'
term and the potential is due to the fact that our `time' is actually
a space-like, radial coordinate. The associated action and Lagrangian
are given by 
\begin{equation}
\label{1deffective_action}
\tilde{S}[\sigma]_{(0,1)} = \frac{1}{2} \int d \tau
(N_{IJ} \dot{\sigma}^I \dot{\sigma}^J + V(\sigma)_{\tilde{q}}) \;.
\end{equation}
Note that this action is {\em not} obtained by 
substituting the definition of the geodesic potential into (\ref{1daction}),
which would lead to a different sign in front of the potential.
Rather, the two Hamiltonians are related through eliminating
the $b^I$ by their equations of motion, and the
associated Lagrangians are in turn given as 
Legendre transforms. This distinction is 
crucial, since 
the elimination of the $b^I$ leads to a non-trivial potential.
To check that the procedure is correct observe that the 
Euler-Lagrange equations of (\ref{1deffective_action})
\[
\ddot{\sigma}^I + \Gamma^I_{JK} \dot{\sigma}^J \dot{\sigma}^K
- \frac{1}{2} N^{IL} \partial_L V (\sigma)_{\tilde{q}} =0 \
\]
agree with (\ref{1dgeodesic_with_potential}), which were  
obtained from the Euler-Lagrange equations (\ref{1dgeodesic})
of the action (\ref{1daction}) by eliminating the $b^I$ through their equation
of motion. 

The problem investigated in \cite{ADOT} is the following:
given an action of the form (\ref{1deffective_action}), 
how can one find new coordinates $\Sigma^I$ and momenta
$P_I$, such that the new momenta are conserved? By 
Hamilton-Jacobi theory this can be achieved by finding a suitable
generating function $\tilde{W}(\sigma, P, \tau)$ of the old coordinates and
new momenta. This function must in particular satisfy
$p_I = \frac{\partial \tilde{W}}{\partial \sigma^I}$
and $\Sigma^I = \frac{\partial \tilde{W}}{\partial P^I}$. 
Since $p_I = N_{IJ} \dot{\sigma}^J$ this leads to a first
order gradient flow driven by the generating $\tilde{W}$:  
$\dot{\sigma}^I = N^{IJ} \partial_J \tilde{W}$ \cite{ADOT}.
For extremal black holes the generating function is independent of 
`time' $\tau$
\cite{ADOT}. 

As we have seen above, the coordinates $\sigma^I$ which we
use throughout this paper have associated canonical momenta $p_I$ 
which are proportional to the charges and hence conserved. This is
due to the extremal instanton ansatz, which solves the constraint
${\cal H}=0$ by imposing that $\dot{\sigma}^I$ and $\dot{b}^I$ 
are proportional up to the constant matrix $R^{I}_{\;\;J}$. Since the
momenta associated with the $b^I$ are conserved as a consequence
of the shift symmetries, the extremal instanton ansatz implies that
the $p_I$ are conserved as well. Above we derived gradient flow equations
(\ref{1storder_gradient_flow}) which are driven by the  `superpotential'
$W = q_I \sigma^I$. As is easily verified, we can interprete
this function as the generating function $\tilde{W} = P_I \sigma^I$ 
of the trivial canonical transformation $\Sigma^I = \sigma^I$, $P_I = p_I$. 
The triviality of the Hamiltonian-Jacobi problem reflects that 
we are already working, for extremal black holes, in the coordinate
system adapted to the symmetries. Note that this does not require
any assumption on the geometry of $M_r$, because for spherically 
symmetric black holes the integrability condition does not impose
constraints on the scalar geometry. 

In the case where the manifold $M_r$ is Hessian, we can go to dual
coordinates 
$\tilde{\sigma}_I \simeq \frac{\partial {\cal V}}{\partial \sigma^I}$
and the momenta are given by $p_I = \dot{\sigma}_I$. 
This observation should be useful when investigating non-extremal black hole
solutions, where the constraint is deformed into ${\cal H} = c^2$. 
We leave a detailed investigation of non-extremal solutions
to future work.

\section{The dual picture}

Given that we interpret the solutions we have constructed as
instantons, we should expect that by substitution of the solution
into the action we obtain a finite and positive result which is
proportional to the instanton charges. But since the scalar
fields vary along null directions of the target 
space it is clear that instanton action, when computed using 
(\ref{paraH-real}) is identically zero. Thus the same feature
which allows to have non-trivial instanton solutions renders
their interpretation as instantons problematic. This is one
aspect of the more fundamental problem of working with a
Euclidean action which is not positive definite. 

The same observations and questions apply to the 
type-IIB D-instanton solution \cite{GGP}
and other stringy instantons,
such as instanton solutions for four-dimensional hypermultiplets
\cite{BGLMM,TheVan,DdVTV,deVVAn,ChiGut}. 
For the purpose of generating higher-dimensional stationary solutions
none of the above points is critical, except perhaps, that
one might expect the ADM mass of black hole or other soliton 
to be related to the action of the instanton obtained
by dimensional reduction 
with respect to time. For the D-instanton and various other
similar instanton solutions it is known that one can obtain 
an instanton action which is
finite, positive and proportional to the instanton charges
by working with a dual version of the action 
(\ref{paraH-real}), which is obtained by dualizing the axionic scalars
$b^I$ into tensor fields. Alternatively, a specific boundary term 
can be added to (\ref{paraH-real}). In this section we derive 
the relevant formulae for the dualization of sigma models of the
type (\ref{paraH-real}). Later we will show that the resulting
instanton actions agree with the masses of the solitons obtained
by dimensional lifting.

In the sigma model (\ref{paraH-real}), the axionic
scalars $b^I$ enter into the field equations only through 
their `field strength' $F_m^I=\partial_m b^I$, which 
can re-expressed in terms of the Hodge-dual three-forms
$H_{mnp|I}$.  By construction, the
three-forms will satisfy the Bianchi identities
\begin{equation}
\label{BianchiH}
\partial_{[m}H_{npq]|I} =0 \;,
\end{equation}
and therefore they can be written, at least locally, as the exterior 
derivatives of two-form gauge fields $B_{mn|I}$.
The standard Lagrangian for a theory of scalars $\sigma^I$ and
two-form gauge fields $B_{mn|I}$ takes the form
\begin{equation}
\label{ScalarTensorLagrangian}
{\cal L} = - \frac{1}{2} N_{IJ}(\sigma) - \frac{1}{2 \cdot 3!}
N^{IJ}(\sigma) H_{mnp|I} H_J^{mnp} \;,
\end{equation}
where 
\[
H_{mnp|I} = 3! \partial_{[m} B_{np|I} \;,
\] 
and where $N^{IJ}$ is the inverse of $N_{IJ}$. Our parametrization
anticipates that the dualization of antiysmmetric tensor fields
into axions inverts the coupling matrix.
The Euclidean form of the Lagrangian is obtained by a Wick rotation,
and the resulting Euclidean action
\begin{equation}
\label{ScalarTensorAction}
S_E[\sigma, B] = - \int d^4x {\cal L}
\end{equation}
is positive definite.\footnote{We include an explicit sign in this
definition, so that the Euclidean action is positive definite instead of 
negative definite.} 
We will now show that this action is equivalent to (\ref{paraH-real}), 
in the sense that it gives rise to the same
equations of motion. 
The first step is to promote the Bianchi identity
$\partial_{[m } H_{npq]|I}=0$ to a field equation 
by introducing a Lagrange multiplier term:
\begin{eqnarray}
S &=& \int d^4 x \Big( 
\frac{1}{2} N_{IJ}(\sigma) \partial_m \sigma^I \partial^m \sigma^J
+ \frac{1}{2\cdot 3!} N^{IJ}(\sigma) H_{mnp|I} H^{mnp}_J \nonumber \\
&&+ \lambda b^I \epsilon^{mnpq} \partial_m H_{npq|J} \Big) \;.
\end{eqnarray}
Here $b^I$ is the Lagrange multiplier for the $I$-th Bianchi
identity, and $\lambda$ is a normalisation constant which we
will fix to a convenient value later. Variation of this action
with respect to $H^{mnp}_I$ gives their equations of motion,
which state that $H^{mnp}_I$ and $\partial_m b^I$ are Hodge dual:
\begin{equation}
\label{Hdualb}
H_I^{mnp} = 3! \lambda N_{IJ}(\sigma) \epsilon^{mnpq}\partial_q b^J \;.
\end{equation}
When we substitue this back into the action, we obtain
\begin{eqnarray}
S[\sigma,b] &=& \int d^4 x \left(
\frac{1}{2} N_{IJ} (\sigma)\partial_m \sigma^I 
\partial^m \sigma^J - \frac{1}{2} (3!\lambda)^2 
N_{IJ} \partial_m b^I \partial^m 
b^J \right) \nonumber \\
&+& (3!\lambda)^2 \oint d^3 \Sigma^m b^I N_{IJ} \partial_m b^J \;.
\end{eqnarray}
The boundary term in the second line results from an integration 
by parts.  We observe that the bulk term 
matches with (\ref{paraH-real}) if we choose $(3!\lambda)^2 =1$. 

The equations of motion for $\sigma^I$ and $B_{mn|I}$ are obtained
by variation of (\ref{ScalarTensorLagrangian}):
\begin{eqnarray}
\partial^m \left( N_{IJ} \partial_m \sigma^J\right) &=&
\frac{1}{2} \partial_I N_{JK} \partial_m \sigma^J \partial^m \sigma^K 
\nonumber \\
\partial^m \left( N^{IJ} H_{mnp|J} \right) &=& 0 \;.
\label{ScalarTensorEOM}
\end{eqnarray}
By construction, they are converted into (\ref{FullEOM})
by substituting in equation (\ref{Hdualb}). 

The action (\ref{ScalarTensorAction}) is positive definite,
and we can obtain instanton solutions by applying 
the Bogomol'nyi trick, i.e. by rewriting the action as
sum of perfect squares, plus a remainder:
\begin{eqnarray}
S[\sigma, B] &=& \int d^4 x \Big[
\frac{1}{2} \left( \partial_m \sigma^I \mp \frac{1}{3!}
N^{IJ} \epsilon_{mnpq} H^{npq}_J \right)^2  \nonumber \\
&& \pm \frac{1}{3!} \partial_m \sigma^I \epsilon^{mnpq} H_{npq|I} \Big] \;.
\end{eqnarray}
Note that the last term is a total derivative as a consequence of the 
Bianchi identity for $H_{mnp|I}$. In contrast to the similar rewriting
(\ref{BogoVar})
used in the previous section, this 
bulk term is not just an alternating sum of squares, but
a single perfect square. Therefore equating the square to zero 
does not just give a saddle point, but a
minimum of the action. 
The resulting equation
\begin{equation}
\label{DualInstantonAnsatz}
\partial_m \sigma^I = \pm \frac{1}{3!} N^{IJ} \epsilon_{mnpq}
H_J^{npq} \;,
\end{equation}
is the Hodge dual version of the extremal 
instanton ansatz (\ref{InstantonAnsatz}), 
as we see immediately using (\ref{Hdualb}).
If the scalar metric $N_{IJ}$ admits a non-trivial $R$-matrix 
(\ref{RotIsometry}), we can impose a Hodge dual version of the
generalized instanton ansatz (\ref{GenInstantonAnsatz}).
As soon as we impose the (generalized) extremal instanton ansatz,
the equations of motion (\ref{ScalarTensorEOM}) reduce
to (\ref{ReducedEOM}). Note that the dual instanton 
ansatz
in combination with the Bianchi identity (\ref{BianchiH})
already implies the equations of motion (\ref{ReducedEOM}).

The extremal instanton ansatz is similar to the (anti-)selfduality condition
characteristic for Yang-Mills instantons. 
This interesting feature is less
obvious when working with the purely scalar version (\ref{paraH-real}) of 
the theory. 
To compute the instanton action, 
we substitute the relation (\ref{DualInstantonAnsatz})
back into the action and obtain:
\begin{equation}
\label{InstAction}
S_{\rm inst} = \int d^4x N_{IJ} \partial_m \sigma^I \partial^m \sigma^J \;.
\end{equation}
This is a boundary term, up to terms proportional to the
equations of motion:
\begin{equation}
S_{\rm inst} = \oint d^3 \Sigma^m N_{IJ} \sigma^I \partial_m \sigma^J \;.
\end{equation}
Guided by the analogy to Yang-Mills instantons, we expect
that this can be expressed in terms of charges. The $B$-field
has an abelian gauge symmetry, $B_{mn} \rightarrow B_{mn}
+ 2 \partial_{[m} \Lambda_{n]}$, and one can define the associated
electric and magnetic charges. For us the magnetic charges
\[
Q_I =  \frac{1}{3!} \oint d^3 \Sigma^m \epsilon_{mnpq}
H^{npq}_I 
\]
will be relevant. The normalization has been chosen such that
they agree with the axionic charges (\ref{DefInstCharge})
when we substitute (\ref{Hdualb}). When evaluated on instanton solutions
these charge take the form
\[
Q_I =  \oint d^3 \Sigma^m N_{IJ} \partial_m \sigma^J \;.
\]
Comparing this to the instanton action, we see that 
the instanton action takes the form
\begin{equation}
\label{InstActChg}
S_{\rm Inst} =  \sigma^I(\infty) Q_I \;,
\end{equation}
provided that the boundary terms corresponding to the localized
$(-1)$-branes (i.e. the centers of the harmonic functions)  
do not contribute. We will investigate this assumption below. 

The boundary term obtained by dualizing the $B$-fields
into axions $b^I$ is
\begin{equation}
\label{BoundaryAction}
S_{\rm bd} = \oint d^3 \Sigma^m b^I N_{IJ} \partial_m b^J =
b^I(\infty) \tilde{Q}_I \;,
\end{equation}
where $\tilde{Q}_I = R_I^{\;\;J}Q_J$ and $\partial_m b^I = R^I_{\;\;J}
\partial_m \sigma^J$. Thus the boundary action equals the instanton
action when evaluated on instanton solutions, provided 
that $b^I(\infty) = R^I_{\;\;J} \sigma^J(\infty)$. Since the $b^I$ are
only defined up to constant shifts, we can regards this as a choice of
gauge. This observation suggests to add the boundary action 
to the scalar bulk action (\ref{paraH-real}), so that by evaluation
on instanton solutions we obtain the same numerical values as for
the scalar-tensor action.

Above we have made the assumption that the instanton solution
is regular at the centers, and that the centers do not contribute 
to the instanton action. Hoewever, the contribution of a single center to the
instanton action is 
\[
\lim_{r\rightarrow 0} \oint_{S_r^3} d^3 \Sigma^m N_{IJ} \sigma^I 
\partial_m \sigma^J =
\lim_{r\rightarrow 0} 2 \pi^2 r^3 N_{IJ} \sigma^I \partial_r \sigma^J \;.
\]
Since $N_{IJ} \partial_m \sigma^J$ is the derivative of a harmonic
function, we know that close to a center
\[
N_{IJ} \partial_r \sigma^J \sim \frac{1}{r^3}  \;.
\]
To have a finite contribution to the instanton action we must 
require that the scalars  $\sigma^I$ have finite limits at the
centers. To obtain (\ref{InstActChg}) we need to impose the
stronger condition that the scalars $\sigma^I$ vanish at the
centers. The standard example of scalar instantons which we have
in mind, including the D-instanton, have this property. Moroever,
for supersymmetric models we expect a relation of the form
(\ref{InstActChg}) between the instanton action and a central
charge of the superysmmetry algebra. Therefore we require
that instanton solutions satisfy
(\ref{InstActChg}). Solutions which do not satisfy this condition
should not be interpreted as proper instantons.

\section{Finiteness of the instanton action and attractor
behaviour.  Examples}

In this section we investigate the behaviour of instanton solutions.
Our main interest is to find criteria which allow us to decide whether
a given Hesse potential allows solutions with finite instanton action 
or not. This requires to investigate the behaviour of solutions at
the centers, which in turn tells us whether solutions exhibit attractor
behaviour, meaning that the asymptotics at the centers is determined
exclusively by the charges, and in particular is independent of the
boundary condition imposed at infinity. The fixed point behaviour 
of extremal black holes is a prototypical example, but we will encounter
a slightly different behaviour which loosely speaking corresponds
to fixed points `at infinity'. Later we will see that for those 
solutions that lift to five-dimensional black holes this behaviour
is nevertheless equivalent to the (five-dimensional) black hole
attractor mechanism. In order to obtain these results, we will need
to make some assumptions about the Hesse potential in order to be able 
to control the asymptotic behaviour of solutions at the centers. 
Two types of Hesse potentials allow a complete analysis: homogeneous
functions and logarithms of homogeneous functions. The second class
corresponds to models which can be lifted to five dimensions in 
presence of gravity. We will also use this section to present a 
variety of explict solutions.

\subsection{Hesse potential ${\cal V}= \sigma^{p}$}

We start with models where the Hesse potential depends on one
single scalar $\sigma$ and is homogeneous of degree $p=N+2$, i.e.
${\cal V} \sim \sigma^{N+2}$. 
Then the metric is proportional to $\sigma^N$, and the sigma model
takes the form
\[
S= \frac{1}{2} \int d^4 x \sigma^N (\partial_m \sigma \partial^m
\sigma - \partial_m b \partial^m b) \;.
\]
The case $N=0, p=2$ corresponds to a free theory. The case $N=3, p=1$
corresponds to Euclidean vector multiplets, obtained by temporal
reduction of five-dimensional vector multiplets. We would like to
include the case $N=-2$, which will turn out to be related to 
supergravity and, more generally, to models including gravity.
Here the Hesse potential is not a homogeneous polynomial, but logarithmic,
${\cal V}=-\log \sigma$.\footnote{Logarithmic Hesse potentials will
be investigated in detail in Sections 4.6 -- 4.8. In Section 6 
we will present a modified formulation of the Hessian geometry of the 
target space,  which is more suitable for this case.}
For $N=-1$ the Hesse potential is the integral 
of the logarithm.

By imposing the extremal instanton ansatz $\partial_m \sigma = \pm 
\partial_m b$, the equation of motion reduces to
\[
\partial_m ( \sigma^N \partial^m \sigma) = 0
\]
which is equivalent to 
\[
\Delta  \sigma^{N+1} = 0  \;.
\]
In other words, $\sigma^{N+1}$ is the dual coordinate of $\sigma$,
which is of course a special case of the relation (\ref{DualCoordinate}).
Close to a center, the solution has the asymptotic form
\[
\sigma^{N+1} \sim \frac{1}{r^2} \;,
\]
which implies that
\[
\sigma \sim  r^{\frac{-2}{N+1}} \;.
\]
Consequently 
\[
\begin{CD}
\sigma @>>{r \rightarrow 0}>
\left\{
\begin{array}{ll}
0 & \mbox{if  } N<-1 \;,\\
\infty &\mbox{if  } N>-1 \;. \\
\end{array} 
\right.
\end{CD}
\]
Therefore a finite action of the form 
$S_{\rm Inst}=\sigma^I(\infty) Q_I$
is obtained for $N=-2,-3, \ldots$, i.e. for logarithmic 
prepotentials and 
for prepotentials which are homogeneous of negative 
degrees $p=-1,-2,\ldots$. 
For models with  $N=0,1,2,3, \ldots$  (i.e. with prepotentials
homogeneous of degree $p=2,3,\ldots$), 
the instanton action is infinite, 
due to contributions from the centers. Therefore these models do not 
possess proper (finite action) instanton
solutions. This includes the case $N=1,p=3$, which corresponds to the 
temporal reduction of five-dimensional vector multiplets.
The case $N=-1$, which is not covered by the above
analysis, has to be treated separately. One finds that $\log \sigma$ 
is harmonic, and therefore the limit at a center is either zero or infinite,
depending on the sign of the charge.

\subsection{General homogeneous Hesse potentials}

We now turn to Hesse potentials which depend on an arbitrary 
number of scalar fields, and are homogeneous of degree $p$.
In this case the dual scalars
\[
\sigma_I \simeq \cal{V}_I =\frac{\partial {\cal V}}{\partial \sigma^I} 
\]
are homogenous functions of degree $p-1$ of the scalars $\sigma^I$. Since
$\Delta \sigma_I =0$, the dual scalars have the 
asymptotics $\sigma_I \sim r^{-2}$ at the centers, implying that
\[
\sigma^I \sim r^{-2/(p-1)} \;.
\]
This is the natural generalization of the result obtained in the
case of a single scalar:
instanton solutions have a finite action of the form 
(\ref{InstActChg}), if the Hesse potential is homogeneous
of degree $p\leq -1$. We will come back to the case of logarithmic
prepotentials later.

As the scalar fields always run off to either 0 or $\infty$ at
the centers,
we need to investigate whether 
these points are at 
finite or infinite `distance'. Since the scalar fields vary
along isotropic submanifolds, the concept of distance has
to be replaced by the concept of an affine curve parameter.
It is sufficient to consider single-centered solutions, and
therefore we have to investigate whether the point $r=0$ is at 
finite or infinite value of an affine parameter along the null
geodesic corresponding to the solution. 
In terms of the dual
scalars the equation of motion is always $\Delta \sigma_I=0$, which, for 
single centered solutions, is the geodesic equation for 
a curve, with the radial variable $r$ as curve parameter:
\[
\Delta \sigma_I = \frac{\partial^2 \sigma_I}{\partial r^2} + 
\frac{3}{r} \frac{\partial \sigma_I}{\partial r} = 0 \;.
\]
Passing to an affine curve parameter
\[
\tau = \frac{A}{r^2} + B 
\]
where $A \not= 0$ and $B$ are constants, we obtain the affine 
version of the geodesic equation. Irrespective of the choice
of affine parameter, we find that 
\[
\begin{CD}
\lim
\tau(r) @>>{r\rightarrow 0}> \infty \;,
\end{CD}
\]
which shows that the point $r=0$ is at infinite affine parameter. 
Therefore the scalars always run away to limit points at `infinite
distance' on the scalar manifold. This is different from the
fixed point behaviour observed for extremal black holes, where
the scalars approach interior points of the scalar manifold,
which are determined by the charges through the black hole
attractor equations. However, for homogeneous prepotentials the run-away
behaviour is not generic and shows features resembling fixed
point behaviour. If we consider ratios
of scalar fields, then the limits at the centers are finite and
depend only on the charges
\[
\frac{\sigma_I}{\sigma_J} \rightarrow \frac{q_I}{q_J} \;.
\]
Thus at least the ratios show fixed point behaviour. 
 
The asymptotic behaviour of the scalars at the centers can 
be represented alternatively by performing a (singular) rescaling,
which brings the limit points to finite parameter values. 
One possible rescaling is to simply rescale the scalars according to
\[
\hat{\sigma}_I := r^2 \sigma_I \;.
\]
Then the new scalar $\hat{\sigma}_I$ 
show proper fixed point behaviour $\hat{\sigma}_I \rightarrow
q_I$. A more intrinsic way of performing a rescaling is to
divide the dual scalars $\sigma_I$ by a homogeneous function
of the scalars, which is chosen such 
that the new scalar fields are homogeneous of degree zero. The 
natural way of achieving this is to take the appropriate power
of the Hesse potential: 
\[
\begin{CD}
\tilde{\sigma}_I  = \frac{\sigma_I}{{\cal V}(\sigma)^{(p-1)/p}} 
@>>{r \rightarrow 0}>
\mbox{finite}\;,
\end{CD}
\]
because
\[
\sigma_I \sim \frac{1}{r^2} \;,\;\;\;
\sigma^I \sim \left(\frac{1}{r^2}\right)^{1/(p-1)} \;,\;\;\;
{\cal V}(\sigma) \sim \left(\frac{1}{r^2}\right)^{p/(p-1)} \;,\;\;\;
{\cal V}(\sigma)^{(p-1)/p}  \sim \frac{1}{r^2}\;.
\]
These rescalings have no immedidate physical meaning, but 
are convenient for visualizing solutions. However, for models with
logarithmic models the rescaling acquires a physical meaning
once we couple the model to gravity, as we will see in Section 6.
Although we did not yet discuss examples with 
logarithmic Hesse potential, it is clear by inspection that 
the above analysis remains valid for the corresponding 
$N=-2$ and $p=0$.

\subsection{Hesse potential  ${\cal V} = \frac{1}{6} C_{IJK} \sigma^I \sigma^J \sigma^K$}

If we construct models by temporal reduction of rigidly 
supersymmetric five-dimensional vector multiplets, then the 
most general Hesse potential is a cubic polynomial \cite{EucI}. Since
constant and linear terms do not enter into the metric, while
quadratic terms only give a constant contribution to the scalar
metric, we can restrict ourselves to homogeneous cubic 
polynomials
\[
{\cal V}(\sigma) = \frac{1}{6} C_{IJK} \sigma^I \sigma^J \sigma^K \;.
\]
The corresponding metric is\footnote{We use a notation where
${\cal V}_I = \frac{\partial {\cal V}}{\partial \sigma^I}$,
${\cal V}_{IJ} = \frac{\partial^2 {\cal V}}{\partial \sigma^I
\partial \sigma^J}$, etc.}
\[
N_{IJ} = {\cal V}_{IJ} = C_{IJK} \sigma^K \;.
\]
The dual coordinates $\sigma_I$, for which the equations of motion
reduce to $\Delta \sigma_I =0$, are normalized according to
\[
\sigma_I = \frac{1}{3} {\cal V}_I=
\frac{1}{6} C_{IJK} \sigma^J \sigma^K = \frac{1}{6}
N_{IJ} \sigma^K  \;.
\]
With this normalization
\[
\sigma_I \sigma^I = {\cal V}(\sigma) \;.
\]
In terms of the dual coordinates, single and multi-centered solutions 
take the form
\[
\sigma_I = h_I + \frac{q_I}{r^2} 
\]
and 
\[
\sigma_I = h_I + \sum_{a=1}^n \frac{q_{Ia}}{|x-x_a|^2}
\]
respectively. In general, we cannot find an explicit expression for
$\sigma^I$ in terms of $\sigma_I$ and, hence, in terms of the 
harmonic functions. 

\subsubsection*{Hesse potential ${\cal V}=\sigma^1 \sigma^2 \sigma^3$}

We now consider a special case where one can obtain explicit 
expressions for the $\sigma^I$. This model is closely related to 
the so-called STU-model. The Hesse potential is
\[
{\cal V} = \sigma^1 \sigma^2 \sigma^3 \;,
\]
and the dual coordinates are chosen\footnote{For convenience,
we have changed the normalization of the $\sigma_I$ compared
to the case of a general cubic Hesse potential.} 
\[
\sigma_1 = \sigma^2 \sigma^3 \;,\;\;\;
\sigma_2 = \sigma^3 \sigma^1 \;,\;\;\;
\sigma_3 =  \sigma^1 \sigma^2 \;.
\]
In terms of dual coordinates, the solution is 
\[
\sigma_I = H_I
\]
where $H_I$, $I=1,2,3$ are harmonic functions. In this case
we can solve explicitly for the $\sigma^I$:
\[
\sigma^1 = \sqrt{\frac{\sigma_2 \sigma_3}{\sigma_1}} =
\sqrt{\frac{H_2 H_3}{H_1}} \;,
\]
with similar expressions for $\sigma^2, \sigma^3$ obtained by cyclic 
permutations.
Here we see explicitly that the fields $\sigma^I$ diverge like
$\frac{1}{r}$ for $r \rightarrow 0$, while their ratios are finite
and only depend on the charges:
\[
\frac{\sigma^1}{\sigma^2} = \frac{H_2}{H_1} \rightarrow \frac{q_2}{q_1}
\;.
\]

\subsection{Hesse potential ${\cal V} = \frac{1}{4!} C_{IJKL} \sigma^I \sigma^J \sigma^K \sigma^L$}

The next example is similar, but not extendable to a supersymmetric
model. 
We take a general quartic Hesse potential
\[
{\cal V} = \frac{1}{4!} C_{IJKL} \sigma^I \sigma^J \sigma^K \sigma^L\;.
\]
The corresponding sigma model is still para-K\"ahler, but not
special para-K\"ahler because the para-K\"ahler potential does not have
a para-holomorphic prepotential. As a shortcut, we 
observe that the corresponding Euclidean sigma model lifts
to a five-dimensional field theory whose couplings are encoded
by a quartic Hesse potential. However five-dimensional supersymmetry
requires a Hesse potential which is at most cubic. 

The corresponding metric is 
\[
N_{IJ} = \frac{1}{2} C_{IJKL} \sigma^K \sigma^L \;,
\]
and dual coordinates are given by 
\[
\sigma_I = \frac{1}{4!} C_{IJKL} \sigma^J \sigma^K \sigma^L 
= \frac{1}{6} {\cal V}_I \;.
\]
The solution is given in terms of harmonic functions 
by $\sigma_I = H_I$. While we cannot solve for the $\sigma^I$
explicitly, homogenity implies that the $\sigma^I \sim r^{-2/3}$ 
for $r\rightarrow 0$, and that the 
ratios $\frac{\sigma_I}{\sigma_J}$ and $\frac{\sigma^I}{\sigma^J}$
have finite limits.

Explicit solutions can be obtained for sufficiently simple
choices of a quartic Hesse potential, for example
\[
{\cal V} = \sigma^1 \sigma^2 \sigma^3 \sigma^4 \;.
\]
Normalizing the dual coordinates such that 
\[
\sigma_1 = \sigma^2 \sigma^3 \sigma^4 \;,\ldots
\]
the solution is
\[
\sigma^1 = \left( \frac{\sigma_2 \sigma_3 \sigma_4}{(\sigma_1)^2}
\right)^{1/3} = \left( \frac{H_2 H_3 H_4}{H_1^2} \right)^{1/3} 
\;, \ldots
\]
with similar expressions for the other $\sigma^I$ obtained by
cyclic permutations.

\subsection{Hesse potential ${\cal V}=-\log(\sigma)$}

In the following sections we discuss models with logarithmic
Hesse potentials. As we will see in Section 6 these are the
models which can be lifted to five-dimensional Einstein-Maxwell type
theories. We will study some aspects already here, because
these models can as well be lifted to five dimension without
coupling to gravity. 

We start with a logarithmic Hesse potential depending on a 
single scalar,
\[
{\cal V} = - \log \sigma
\]
where $\sigma >0$. The resulting Hessian metric is
\[
{\cal V}'' = \frac{1}{\sigma^2}  \;.
\]
We have already seen that this model is in the class where
the instanton action has the form (\ref{InstActChg}). The dual
coordinate is proportional to ${\cal V}'$, and we normalize it
to be $\frac{1}{\sigma}$. The reduced equation of motion is
\[
\Delta \frac{1}{\sigma} = 0 \;,
\]
which is solved by
\[
\sigma = \frac{1}{H} \;,
\]
where $H$ is a harmonic function. Considering  a single centered solution,
\[
\sigma = \frac{1}{h+\frac{q}{r^2}} \;,
\]
we can see explicitly how $\sigma$ behaves for $r\rightarrow \infty$
and $r\rightarrow 0$:
\[
\begin{CD}
\sigma 
@>>{r \rightarrow \infty}> \frac{1}{h} \;,\;\;\;
\sigma 
@>>{r \rightarrow 0}> 0 \;.
\end{CD}
\]
This illustrates our general result, and we can see explicitly that 
the action is finite.

The target space corresponding to this model is the symmetric
space $SL(2,\mathbbm{R})/SO(1,1)$, which is also known as
$AdS^2$. The action, expressed in terms of the scalars $\sigma$ and
$b$ is
\[
S = \int d^4x \frac{1}{\sigma^2} ( \partial_m \sigma \partial^m \sigma
- \partial_m b \partial^m b) \;.
\]
In terms of the para-complex coordinates
 $X=\sigma + e b$ this becomes
\[
S = \int d^4x \frac{\partial_m X \partial^m \bar{X}}{ (\mbox{Re}(X))^2 } \;,
\]
which makes explicit that the target space is a para-K\"ahler 
manifold with para-K\"ahler potential 
\[
K = - \log (X+\bar{X}) \;.
\]
By the analytic continuation $b \rightarrow ib$ we obtain  
the upper half plane, equipped with the Poincar\'e metric,  
${\cal H} \cong \frac{SL(2,\mathbbm{R})}{SO(2)}$.\footnote{Various
coordinate systems for the two symmetric spaces in question 
can be found, for example, in \cite{Gilmore}.}

\subsection{Hesse potential ${\cal V}=-\log(\sigma^1 \sigma^2 \sigma^3 )$}

Another model, which turns out to be the Euclidean version of
the well-known STU-model is obtained by taking three copies 
of the previous model. The Hesse potential is
\[
{\cal V} = - \log  (\sigma^1 \sigma^2 \sigma^3) =
-\log \sigma^1 -\log \sigma^2 -\log \sigma^3 \;.
\]
The corresponding target space is the product of three copies
of $SL(2,\mathbbm{R})/SO(1,1)$, which is para-K\"ahler with 
para-K\"ahler potential
\[
K = -\log ( (X^1+\bar{X}^1)(X^2+\bar{X}^2)(X^3+\bar{X}^3) \;,
\]
where $X^I = \sigma^I + e b^I$. This target space
is in fact even projective special para-K\"ahler, with para-holomorphic
prepotential $F= - \frac{X^1 X^2 X^3}{X^0}$, as it must be 
for Euclidean vector multiplets coupled to supergravity \cite{EucIII}. 

The dual coordinates can be normalized to be 
\[
\sigma_I =  \frac{1}{\sigma^I} \;,
\]
so that explicit solutions for the $\sigma^I$ can be found: 
\[
\sigma^I = \frac{1}{H_I} \;.
\]
We will see later that this solution can be lifted to a five-dimensional
extremal black hole solution of five-dimensional supergravity. 

\subsection{Hesse potential ${\cal V} = - \log \hat{\cal V}(\sigma)$,
with homogeneous $\hat{\cal V}(\sigma)$}

Finally, we consider the general case of a Hesse potential which 
is the logarithm of a homogeneous function $\hat{\cal V}(\sigma)$
(of arbitrary degree):
\[ 
{\cal V}(\sigma^I) = - \log \hat{\cal V}(\sigma^I)
\]
where
\[
\hat{\cal V}(\lambda \sigma^I) = \lambda^p \hat{\cal V}(\sigma^I) \;
\]
with integer $p$. 
Then the Hesse potential is not strictly a homogeneous function, but
it is homogeneneous of degree zero up to a constant shift. However,
the first derivatives
\[
\sigma_I \simeq \frac{\partial \cal V}{\partial \sigma^I}
\]
are homogeneous of degree $-1$, and the metric, which is 
given by the second derivatives,
\[
N_{IJ} = \frac{\partial^2 \cal V}{\partial \sigma^I \partial \sigma^J}
\]
is homogeneous of degree $-2$. This corresponds to the case $N=-2$
discussed in Sections 4.1 and 4.2, and the results derived there 
apply (setting $N=-2$ and $p=0$ in te relevant formulae. 
In particular the instanton action 
is of the form (\ref{InstActChg}), and the solutions show a version 
of fixed point behaviour where the scalars run off to a point at infinite
affine parameter while the ratios approach finite values
determined by the charges.

\section{Lifting to five dimensions, without gravity}

In the following section we discuss the lifting
of instanton solutions to five-dimensional solitons in 
the absence of gravity. Here no constraints need to be
imposed on the Hesse potential. We show that the mass
of the soliton obtained by lifting is equal to the instanton 
action. 

The para-Hermitean Euclidean sigma model (\ref{paraH-real}) can be lifted 
to a five-dimensional theory of scalars and gauge fields:
\begin{equation}
\label{5dRigidAction}
S[\sigma, A_\mu] 
= \int d^5 x \left( - \frac{1}{2} N_{IJ}(\sigma) \partial_\mu
\sigma^I \partial^\mu \sigma^J - \frac{1}{4} N_{IJ}(\sigma)
F^I_{\mu \nu} F^{\mu \nu|J} + \cdots \right) \;.
\end{equation}
Here $\mu, \nu=0,1,\cdots, 4$ are five-dimensional Lorentz
indices, and the four-dimensional axions have been identified
with the time components of the five-dimensional gauge fields
\[
b^I = - A^I_0 \;.
\]
To obtain a covariant theory, we have added the magnetic
components $F_{mn}^I$, $m,n=1,\ldots,4$ of the five-dimensional
field strength. We also allow further terms, as long as they
do not contribute to the four-dimensional sigma model obtained
by reduction over time.
It is straightforward to verify that the
five-dimensional action (\ref{5dRigidAction}) reduces to the
para-Hermitean sigma model (\ref{paraH-real}) upon restricting to 
static and purely electric field configurations, and reducing
with respect to time. Thus instanton 
solutions of (\ref{paraH-real}) lift to electrically charged solitons 
of (\ref{5dRigidAction}).

The full field equations of the five-dimensional theory 
have the following form. The equation of motion for the
scalars $\sigma^I$ is 
\[
N_{KJ} \Box \sigma^J + \frac{1}{2} \partial_K N_{IJ} 
\partial_\mu \sigma^I \partial^\mu \sigma^J = \frac{1}{4}
\partial_K N_{IJ} F^{I}_{\mu \nu} F^{\mu \nu|J} \;,
\]
and the equation of motion of the five-dimensional gauge fields is
\[
\partial_\mu ( N_{IJ} F^{\mu \nu|J}) =0 \;.
\]
If we impose that the solution is static and does not carry magnetic
charge, then all time-derivatives vanish and the only non-vanishing
field strength components can be expressed in terms of the
electrostatic potentials $A_0^I$:
\[
F_{tm} = - F_{mt} = - \partial_m A_0^I  = \partial_m b^I\;.
\]
In such backgrounds the equations of motion take the
following form:
\begin{eqnarray}
N_{KJ} \Delta \sigma^J + \frac{1}{2} \partial_K N_{IJ} 
\partial_m \sigma^I \partial^m \sigma^J &=& \frac{1}{2}
\partial_K N_{IJ} F^I_{0m} F^{0m|J} 
\nonumber  \\
\partial_m ( N_{IJ} \partial^m A^J_0 ) &=& 0 \;.
\end{eqnarray}
Expressing $F^I_{0m}$ and $A^I_0$ in termes of $b^I$, 
we see that these equations of motion are identical to 
(\ref{FullEOM}).
The extremal instanton ansatz corresponds to imposing
\[
\partial_m \sigma^I = \pm F_{0m}^I 
\]
which means that the scalars $\sigma^I$ are proportional to the 
electrostatic potentials. For five-dimensional vector multiplets, this
is the condition for a BPS solution supported by scalars and electric
fields. 

Imposing the extremal instanton ansatz we therefore obtain the reduced
equations of motion
\[
\partial_m (N_{IJ} \partial^m \sigma^J ) =0 \;,
\]
which is identical to (\ref{ReducedEOM}).
The four-dimensional instanton charges equal
the five-dimensional electric charges, which are defined by
\[
Q_I = \oint d^3 \Sigma^m N_{IJ} F^J_{0m} = \oint d^3 \Sigma^m
N_{IJ} \partial_m b^J  \;.
\]
From the five-dimensional point of view the method used in 
Section 2.3 to solve the equations of motion is a standard 
method for solving Maxwell-type equations in an electrostatic background.

We expect that the four-dimensional instanton action is 
related to the five-dimensional mass. The mass of the soliton 
is obtained by integrating the energy density, which is 
the component $T_{00}$ of the energy momentum tensor 
$T_{\mu \nu}$, over space. We use the symmetric energy momentum
tensor which is obtained by coupling the action (\ref{5dRigidAction})
to a background metric and varying it. The result is 
\begin{eqnarray}
T_{\mu \nu} &=& N_{IJ} \partial_\mu \sigma^I \partial_\nu \sigma^J
- \frac{1}{2} N_{IJ} \eta_{\mu \nu} \partial_\rho \sigma^I 
\partial^\rho \sigma^J \nonumber \\
&&+ N_{IJ} F^I_{\mu \rho} F_{\nu}^{\;\;\rho|J} - \frac{1}{4} \eta_{\mu \nu}
F^I_{\rho \sigma} F^{\rho \sigma|J} \;. 
\end{eqnarray}
In a static, purely electric background, the resulting energy density
is 
\[
T_{00} = \frac{1}{2} N_{IJ} \partial_m \sigma^I \partial^m \sigma^J
+ \frac{1}{2} N_{IJ} \delta^{mn} F_{0m}^I F_{0n}^J  \;.
\]
For solutions where $F^I_{0m}=\pm \partial_m \sigma^I$,
this becomes
\[
T_{00} = N_{IJ} \partial_m \sigma^I \partial^m \sigma^J \;.
\]
The integral expression for the soliton 
mass $M$ of the soliton agrees with the instanton 
action (\ref{InstAction}) and the boundary action (\ref{BoundaryAction}):
\[
M=\int d^4 x T_{00} = S_{\rm inst.} = S_{\rm bound.}  \;.
\]
Our previous discussion of fixed point behaviour of the scalars $\sigma^I$
remains valid, because there is no difference between the four- and
five-dimensional scalars. In particular the soliton mass is
finite if the $\sigma^I$ approach finite values, and it is 
given in terms of the five-dimensional electric charges 
as 
\[
M=\sigma^I(\infty) Q_I \;,
\]
if the scalars go to zero at the centers. Models where the 
Hesse potential is homogeneous of positive degree do not have
proper, i.e. finite mass, solitons of the type considered. This
includes rigid five-dimensional vector multiplets. If the degree
of homogeneity is negative, or if the Hesse potential is the 
logarithm of a homogeneous function, then solitons with finite mass
do exist.

\section{Lifting to five dimensions, with gravity}

\subsection{Dimensional lifting and dimensional reduction}

We now turn to the lifting of four-dimensional instantons to
five-dimensional black holes. In the presence of gravity the
relation between the five-dimensional and four-dimensional actions
becomes more complicated. As a first step we would like to identify the class
of five-dimensional Einstein-Maxwell type actions which 
reduce to actions of the form (\ref{SigmaR}) by dimensional
reduction with respect to time. To be precise we allow 
additional terms in both actions, as long as 
(\ref{SigmaR}) is a consistent reduction, i.e., as long
as solutions of (\ref{SigmaR}) are solutions of the five-dimensional
theory. The main new feature in the presence of gravity 
is that the decomposition of the
five-dimensional metric gives rise to a Kaluza-Klein scalar
and Kaluza-Klein gauge field. The Kaluza-Klein gauge field
can be set to zero consistently. At the level of five-dimensional
solutions this means that we restrict ourselves to solutions which 
are not only stationary, but static, i.e. we exclude rotating solutions. 
However the Kaluza-Klein scalar provides a complication because 
it needs to be incorporated into the four-dimensional scalar 
sigma model.\footnote{As will be explicit from the solutions discussed
later, freezing the Kaluza-Klein scalar is not an option, since this
would only leave us with trivial solutions.}

One class of examples where one obtains Euclidean actions of the
type (\ref{SigmaR}) is the temporal reduction of five-dimensional 
supergravity coupled to vector multiplets \cite{EucIII}. We will
adopt the strategy of generalizing this class while keeping the 
relevant feature that temporal reduction gives rise to a para-K\"ahler
sigma model. As far as the 
relation between five-dimensional and four-dimensional
actions is concerned, the analysis can be carried out in parallel for spatial
and temporal reduction. For concreteness we will take the case
of temporal reduction, but the other case is simply obtained by
flipping signs in the Lagrangian, as discussed in more detail in 
\cite{EucIII}. 

The geometry underlying five-dimensional supergravity with vector 
multiplets is the local (or projective) version of very special
real geometry \cite{GST}. This is a type of Hessian 
geometry, where the Hesse potential ${\cal V} = - \log \hat{\cal V}$
is the logarithm of a 
homogeneous cubic polynomial $\hat{\cal V}$, which is called the
prepotential. To be precise, the metric obtained from this 
Hesse potential gives the coupling matrix of the gauge fields, 
while the scalar metric is its pull back to the hypersurface
$\hat{\cal V}=1$. This reflects that the supergravity theory 
has one scalar field less than it has gauge fields. A five-dimensional
vector multiplet contains a gauge field and a real scalar, but
the gravity multiplet contains an additional gauge field,
the graviphoton. Upon dimensional reduction each gauge field gives
rise to an axionic scalar, which can be combined with 
the five-dimensional scalars and the Kaluza-Klein scalar into
a sigma model of the type (\ref{SigmaR}). Thus it is important
to have one additional gauge field in five dimensions. The other
critical feature is that the metric of the five-dimensional scalar
sigma model is homogeneous of degree $-2$ in the scalar fields. 
As we will see later
this is crucial for combining the Kaluza-Klein scalar with 
the five-dimensional scalars in such a way that we obtain a 
sigma model of the form (\ref{SigmaR}). 

As we have seen in Section 4, a Hessian metric is homogeneous of
degree $-2$ if its Hesse potential ${\cal V}=-\log \hat{\cal V}$
is the logarithm of 
a homogeneous function $\hat{\cal V}$, irrespective of the
degree of homogenity. Therefore we will generalize the
local very special real geometry of supergravity by dropping
the requirement that the prepotential $\hat{\cal V}$ is a homogeneous 
cubic polynomial, while still requiring that it is a homogeneous
function of degree $p$, where $p$ is now arbitrary. 

Dimensional reduction of five-dimensional supergravity with
vector multiplets with respect to space results in target space
geometries which are projective special K\"ahler \cite{GST}. 
The map between
the target geometries of five-dimensional and four-dimensional
vector multiplets is the $r$-map 
\cite{dWvP:1992}, which we will call the local (or projective)
$r$-map, to disinguish it from its rigid (or global) counterpart. 
If one reduces over time, one obtains projective
special para-K\"ahler manifolds, and the corresponding map 
is called the local (or projective) para-$r$-map \cite{EucIII}.
The following construction provides a generalization of both 
the local $r$-map and local para-$r$-map. For concreteness
we will give explicit expression for the para-$r$-map, 
and explain in the end how the $r$-map is obtained by analytical
continuation.

The construction starts with 
$n+1$ scalar fields
$h=(h^I)=(h^0,h^1, \ldots, h^n)$, which we interprete as
affine coordinates on an $(n+1)$-dimensional Hessian manifold
$\tilde{M}_r$. We work locally and take $\tilde{M}_r$ to be
an open domain in $\mathbbm{R}^{n+1}$.
The Hesse potential for this manifold (which will
be the prepotential for the actual scalar manifold $M_r$)
is ${\cal V}(h) = - \log {\hat {\cal V}}(h)$, where 
the prepotential $\hat{\cal V}(h)$ is homogeneous of degree $p$:
\begin{equation}
\hat{\mathcal{V}}(\lambda h^0,..,\lambda h^n)=
\lambda^p \hat{\mathcal{V}}(h^0,h^1,\ldots, h^n)\;. 
\end{equation}
Taking the derivative with respect to $\lambda$ we obtain
\begin{eqnarray}
\hat{\mathcal{V}}_I(\lambda h)h^I=p\lambda^{p-1}\hat{\mathcal{V}}(h) \;,
\end{eqnarray}
where the subscript $I$ denotes differentiation with respect to $h^I$. 
By setting $\lambda=1$ we obtain 
\begin{equation}
\hat{\mathcal{V}}_Ih^I=p\hat{\mathcal{V}}(h) \;.\label{Euler}
\end{equation}
Further differentiation implies that 
\begin{equation}
\hat{\mathcal{V}}_{IJ}h^I=(p-1)\hat{\mathcal{V}}_J  \label{Euler2}\;.
\end{equation}
The logarithm of $\hat{\cal V}(h)$ is used to define a Hessian metric 
by
\[
a_{IJ}(h) = - \frac{1}{p}
\frac{\partial^2 \log \hat{\cal V}(h)}{\partial h^I \partial h^J}
= - \frac{1}{p} \left( \frac{ \hat{\cal V}_{IJ}}{\hat{\cal V}} 
- \frac{ \hat{\cal V}_I \hat{\cal V}_J}{ \hat{\cal V}^2} \right)\;.
\]
A conventional factor $\frac{1}{p}$ has been introduced in
order to be consistent with supergravity conventions for $p=3$.
The metric is homogeneous of degree $-2$ in the $h^I$. In order
to ensure that the metric $a_{IJ}(h)$ is positive definite, we
might need to restrict the   fields $h=(h^I)$ 
to a suitable domain $D \subset \mathbbm{R}^{n+1}$.
The scalar target manifold $M_r$ of the model is 
the hypersurface $\{h^I | \hat{\cal V}(h) =1\}$ of $D$, equipped with the
pull-back metric. 
\[
a_{xy}(\phi) = \frac{\partial h^I}{\partial \phi^x} 
\frac{\partial h^J}{\partial \phi^y} a_{IJ}(h(\phi)) \;.
\]
The physical scalars $\phi^x$, $x=1,\ldots, n$ provide 
local coordinates on the hypersurface 
$\{h^I|  \hat{\cal V}=1 \} \subset D$.

In the following it will be convenient to work with 
the fields $h^I$, which are subject to the constraint $\hat{\cal V}(h)=1$,
and with the associated Hessian metric $a_{IJ}(h)$. We will need a few
relations involving $a_{IJ}(h)$. First note
that (\ref{Euler}) and (\ref{Euler2}) can be used to show that
\begin{eqnarray}
a_{IJ}(h)h^Ih^J&=&
- \frac{1}{p}\partial_I\partial_J\log \hat{\mathcal{V}}(h)h^Ih^J 
= - \frac{1}{p}\left(\frac{\hat{\mathcal{V}}_{IJ}}{\hat{\mathcal{V}}} 
- \frac{\hat{\mathcal{V}}_I \hat{\mathcal{V}}_J}
{\hat{\mathcal{V}}^2}\right)h^Ih^J \nonumber \\ 
&=& \frac{ \hat{\mathcal{V}}_Jh^J}{p\hat{\mathcal{V}}} =1 \;. 
\label{eq:Vrelation} 
\end{eqnarray}
Differentiation of the constraint $\hat{\cal V}(h)=1$ with respect to
space-time implies
\begin{equation}
\hat{\mathcal{V}}_I\partial_\mu h^I = 0 \;,
\end{equation}
where $\mu=0,\ldots, 4$ are five-dimensional space-time indices.
Combining this with (\ref{eq:Vrelation}) we obtain
\begin{equation}
a_{IJ}h^I\partial_\mu h^J= -\frac{\hat{\mathcal{V}}_J}{\hat{\mathcal{V}}}
\partial_\mu h^J=0 \;. \label{eq:Vrelation2}
\end{equation}

We now use the prepotential $\hat{\cal V}(h)$ to define the following 
five-dimensional bosonic Lagrangian:
\begin{equation}
\label{5dLagrangian}
\hat{e^{-1}}\hat{\mathcal{L}}= \frac{\hat{R}}{2}-\frac{3}{4}a_{IJ}(h)
\partial_\mu h^I\partial^\mu h^J - \frac{1}{4}a_{IJ}(h) F_{\mu\nu}^I 
F^{\mu\nu J} + \cdots \;.
\end{equation}
Here $\hat{R}$ is the five-dimensional Ricci scalar, 
$\hat{e}$ is the determinant of the local frame (`f\"unfbein'),
$a_{IJ}(h)$ is
the Hessian metric defined above, and for the scalar term the
constraint $\hat{\cal V}(h)=1$ is understood. As indicated, the 
Lagrangian might contain further terms, provided that these do not
contribute to four-dimensional Euclidean sigma model obtained by 
reduction with respect to time. For $p=3$, (\ref{5dLagrangian}) 
is part of the Lagrangian of five-dimensional vector multiplets coupled
to $n$ vector multiplets. The full supergravity Lagrangian also contains
a Chern-Simon terms and fermionic terms, which, however, do not 
contribute to the four-dimensional sigma model upon reduction.

We now reduce the Lagrangian (\ref{5dLagrangian}) with respect to 
time.\footnote{We refer to \cite{EucIII} for a more detailed discussion 
of dimensional reduction.} The reduction of the metric is carried out in 
such a way
that the resulting four-dimensional Einstein-Hilbert term
has the canonical form, i.e., we reduce from the five-dimensional
Einstein frame to the four-dimensional Einstein frame. The corresponding
parametrization of the line element is
\[
ds^2_{(5)} = - e^{2\tilde{\sigma}} (dt + {\cal A}_m dx^m)^2 +
e^{-\tilde{\sigma}} ds^2_{(4)} \;,
\]
where $\tilde{\sigma}$ is the Kaluza-Klein scalar and ${\cal A}_m$ is the
Kaluza-Klein vector. Upon dimensional reduction over time,
the   zero components $\mathcal{A}^I_0$ of the five-dimensional gauge fields
become  four-dimensional scalar fields $m^I= \mathcal{A}^I_0$. 
In four dimensions, we only keep the Einstein-Hilbert term and the
scalar terms. This is a consistent truncation and corresponds
to the restriction to five-dimensional field configurations 
which are static and purely electric. 
The relevant part of the reduced Lagrangian is 
\begin{equation}
e^{-1} \mathcal{L}=\frac{R}{2}-\frac{3}{4}\partial_m\tilde{\sigma}
\partial^m\tilde{\sigma} 
- \frac{3}{4}a_{IJ}(h)\partial_m h^I \partial^m h^J + \frac{1}{2}
e^{-2\tilde{\sigma}}
a_{IJ}(h)\partial_m m^I \partial^m m^J \;,
\end{equation}
where $m=1,\ldots, 4$ are indices in four-dimensional space, 
$R$ is the four-dimensional
Ricci scalar, and $e$ is the determinant of the four-dimensional 
local frame (`vierbein').
By making the redefinitions 
\begin{eqnarray}
h^I&=&Ae^{-\tilde{\sigma}}\sigma^I\;, \\
m^I&=&B b^I \;,
\end{eqnarray}
where $A,B$ are constants to be fixed later,
the Lagrangian takes on the form
\begin{eqnarray}
e^{-1}\mathcal{L} &=& \frac{R}{2} 
-\frac{3}{4}\partial_m\tilde{\sigma}\partial^m\tilde{\sigma} 
- \frac{3}{4}a_{IJ}(e^{-\tilde{\sigma}}\sigma)\sigma^I \sigma^J
\partial_m e^{-\tilde{\sigma}}\partial^m 
e^{-\tilde{\sigma}} \nonumber \\
&&  - \frac{3}{4}a_{IJ}(e^{-\tilde{\sigma}}\sigma)e^{-2\tilde{\sigma}}
\partial_m \sigma^I\partial^m \sigma^J 
-\frac{3}{2}a_{IJ}(e^{-\tilde{\sigma}}\sigma^)e^{-\tilde{\sigma}} \sigma^I
\partial_m e^{-\tilde{\sigma}}\partial^m \sigma^J 
\nonumber \\
&& + \frac{B^2}{2A^2}e^{-2\tilde{\sigma}}a_{IJ}(e^{-\tilde{\sigma}} \sigma)
\partial_m b^I\partial^m  b^J \;.
\end{eqnarray}
From now on we regard $\sigma^I$ and $b^I$ as the independent fields. 
Note that the constraint $\hat{\cal V}(h) =1$ implies the
relation 
\begin{equation}
\label{RelKK}
\hat{\cal V}(\sigma) = \hat{\cal V}(A^{-1} e^{\tilde{\sigma}} h) = 
A^{-p} e^{p \tilde{\sigma}} \hat{\cal V}(h) = A^{-p} e^{p \tilde{\sigma}} \;,
\end{equation}
which expresses the Kaluza-Klein scalar $\tilde{\sigma}$
as a function of the four-dimensional scalars $\sigma^I$.

By considering the relations (\ref{eq:Vrelation}) and (\ref{eq:Vrelation2}), 
the first and second term cancel and the fourth term vanishes.  
If we choose the constants $A,B$ to satisfy 
\begin{equation}
B^2=\frac{3A^2}{2} \;,
\end{equation}
and use that $a_{IJ}(h)$ is homogeneous of degree $-2$, 
the remaining terms in the Lagrangian take the form
\begin{eqnarray}
e^{-1}\mathcal{L} &=& \frac{R}{2} - \frac{3}{4} 
a_{IJ}(\sigma) \partial_m \sigma^I \partial^m \sigma^J
+ \frac{3}{4} a_{IJ}(\sigma) \partial_m b^I \partial^m b^J \;.
\end{eqnarray}
Defining
\begin{equation}
N_{IJ}(\sigma)=\frac{3}{2} a_{IJ}(\sigma) \;,
\end{equation}
we recognize the standard form (\ref{SigmaR}) of a 
para-Hermitean sigma model with $n$ commuting shift isometries,
coupled to gravity, 
\begin{equation}
\label{ReducedSigmaTime}
e^{-1}\mathcal{L}= \frac{R}{2} - \frac{1}{2}
N_{IJ}(\sigma)(\partial_m \sigma^I \partial^m \sigma^J 
- \partial_m b^I \partial_m b^J)  \;.
\end{equation} 
The metric $N_{IJ}(\sigma)$ has
the Hesse potential ${\cal V}(\sigma)=-\log \hat{\cal V}(\sigma)$:
\[
N_{IJ}(\sigma) = -\frac{3}{2p} \frac{\partial^2}{\partial
\sigma^I \partial \sigma^J} \log \hat{\cal V}(\sigma) \;.
\]
As a result,
the metric $N_{IJ} \oplus (-N_{IJ})$ of the scalar manifold
spanned by $\sigma^I, b^I$ is para-K\"ahler. 
This is seen explicitly by introducing para-holomorphic coordinates
\[
X^I = \sigma^I + e b^I \;,
\]
and computing
\[
\frac{\partial^2 \log \hat{\cal V}}{\partial X^I \partial \bar{X}^J} =
\frac{\partial^2 \log \hat{\cal V}}{\partial \sigma^K \partial \sigma^L} 
\frac{\partial \sigma^K}{\partial X^I}\frac{\partial \sigma^L}{\partial 
\bar{X}^J} = \frac{1}{4} \frac{\partial^2 \log \hat{\cal V}}{\partial \sigma^I
\partial \sigma^J} = \frac{p}{6} N_{IJ}  \;.
\]
Thus $K(X,\bar{X})= \frac{6}{p} \log \hat{\cal V}$ is a para-K\"ahler 
potential for the metric $N_{IJ}\oplus (-N_{IJ})$.

The relation between the five- and four-dimensional Lagrangian 
is true irrespective of the value of $p$ that we choose, and hence 
it makes sense for models with $p\not=3$, which cannot be embedded into
a five-dimensional supersymmetric model. However, it was crucial
that we could combine the Kaluza-Klein scalar with the five-dimensional
scalars $h^I$ in such a way that the scalar target manifold of the
reduced theory became 
para-Hermitean. This worked only because the metric $a_{IJ}(h)$ is
homogeneous of degree $-2$. Therefore, there is no obvious further
generalization which would allow one to drop the condition that the
prepotential is homogeneous. The effect of reducing over space 
rather than time is to replace 
$b^I$ by $ib^I$ in \ref{ReducedSigmaTime}. Equivalently, in 
terms of (para-)complex coordinates, it corresponds to 
replacing $X^I=\sigma^I + e b^I$ by
$Y^I = \sigma^I + i b^I$, i.e. the para-complex structure is replaced
by a complex structure, and one obtains a K\"ahler manifold where 
the K\"ahler potential is proportional to 
the prepotential $\hat{\cal V}(\sigma)$.
Thus, as in \cite{EucIII} the 
para-$r$-map and $r$-map are related by analytic continuation (see 
also Appendix A).

Having fixed the relation between the five-dimensional and the
four-dimensional theory, we can now see how four-dimensional
instantons lift to five-dimensional solutions. We have
restricted ourselves to solutions of (\ref{SigmaR}) where
the four-dimensional metric is flat, 
$ds^2_{(4)} = \delta_{mn} dx^m dx^n$.
Such line elements lift to five-dimensional line elements 
of the form 
\[
ds^2_{(5)} = - e^{2\tilde{\sigma}} dt^2 + e^{-\tilde{\sigma}} \delta_{mn}
dx^m dx^n \;,
\]
where $\tilde{\sigma}$ is the Kaluza-Klein scalar. This
is precisely the structure of a line element for an extremal 
five-dimensional black hole. Extremal black holes have the particular
feature that their line elements reduce under temporal reduction 
to flat line elements, provided that one uses the Einstein frame
in both dimensions. The non-trivial five-dimensional geometry is 
fully captured by the Kaluza-Klein scalar, while the four-dimensional
metric is flat.
This explains why extremal black holes correspond
to null geodesics, and why we could effectively drop the four-dimensional
Einstein-Hilbert term when constructing solutions. 
This observation provides additional justification for 
calling the corresponding instanton solutions extremal. 

From the four-dimensional point of view all information is encoded
in the scalar fields $\sigma^I$. With the choice $A=1$, which implies 
$B=\sqrt{\frac{3}{2}}$,
the Kaluza-Klein scalar is determined by the four-dimensional 
scalars through the relation
\begin{equation}
\label{RelKK2}
e^{p\tilde{\sigma}} = \hat{\cal V}(\sigma) \;,
\end{equation}
while the five-dimensional scalars are given by 
\[
h^I = e^{-\tilde{\sigma}} \sigma^I  \;,
\]
We have a Hesse potential of the
form ${\cal V}(\sigma) = - \log \hat{\cal V}(\sigma)$, and therefore
the dual scalars have the form 
\[
\sigma_I \simeq \frac{\partial }{\partial \sigma^I} \log \hat{\cal V}(\sigma)\;.
\]
As in previous examples we will fix the factor of proportionality
at our convenience. The solution is given by $\sigma_I(x)=H_I(x)$, where
$H_I(x)$ are harmonic functions on $\mathbbm{R}^4$. 
Explicit expressions for the $\sigma^I$
can only be obtained case by case if the prepotential is sufficiently
simple. However, the asymptotics of the solution at the center is 
known from Section 4, and we will see below that this allows us to 
obtain information  about the ADM mass and about the black hole entropy. 
The axions $b^I$ are determined
by the extremal instanton ansatz and in turn determine the five-dimensional
gauge fields.

\subsection{ADM mass and instanton action}

Before looking into explicit examples, we show that the
ADM mass of the five-dimensional black hole is equal to the
action of the corresponding four-dimensional instanton.
The ADM mass can be written as a surface integral involving
the Kaluza-Klein scalar $\tilde{\sigma}$ \cite{EucIII}. To compare
this to the instanton action, we express the ADM mass in terms of the
prepotential:
\[
M_{ADM} = - \frac{3}{2} \oint d^3 \Sigma^m \partial_m 
e^{-\tilde{\sigma}} = - \frac{3}{2} \oint d^3 \Sigma^m
\partial_m \hat{\cal V}(\sigma)^{-1/p}  \;.
\]
Now we compare this to the instanton action
\[
S_{\rm inst} = \oint d^3 \Sigma^m N_{IJ} \sigma^I \partial_m
\sigma^J  \;.
\]
The metric $N_{IJ}$ is given by
\[
N_{IJ} = - \frac{3}{2p}  \left( \frac{ \hat{\cal V}_{IJ}}
{ \hat{\cal V}}
- \frac{ \hat{\cal V}_I \hat{\cal V}_J }{ \hat{\cal V}^2}  \right) \;.
\]
Using that $\hat{\cal V}(\sigma)$ is homogeneous of degree $p$, we find
\[
N_{IJ} \sigma^I \partial_m \sigma^J 
=
- \frac{3}{2p} \left( 
\frac{ \hat{\cal V}_{IJ} \sigma^I }{ \hat{\cal V} } -
\frac{ \hat{\cal V}_I \sigma_I \hat{\cal V}_J }{ \hat{\cal V}^2 } \right)
\partial_m \sigma^J
=
 \frac{3}{2p}  \frac{ \hat{\cal V}_J }{ \hat{\cal V} } 
\partial_m \sigma^J 
\]
But this is a total derivative:
\[
N_{IJ} \sigma^I \partial_m \sigma^J 
=
 \frac{3}{2p} \partial_m \log \hat{\cal V}(\sigma) \;.
\] 
As a result we have
\begin{eqnarray}
M_{ADM} & =& - \frac{3}{2} \oint d^3 \Sigma^m \partial_m 
\hat{\cal V}(\sigma)^{-1/p}  = - \frac{3}{2} \oint d^3 \Sigma^m
\partial_m e^{-\tilde{\sigma}} \;, 
\nonumber \\
S_{\rm inst} 
&=& \frac{3}{2} 
\oint d^3 \Sigma^m \partial_m \log \hat{\cal V}(\sigma)^{1/p} 
= \frac{3}{2} \oint d^3 \Sigma^m \partial_m \tilde{\sigma} \;.
\end{eqnarray}
Both the ADM mass and the instanton action are surface integrals,
but the integrands are different. To compare the integrals 
we rewrite the ADM mass as
\begin{equation}
\label{ADMvsInst}
M_{ADM} = \frac{3}{2} \oint d^3 \Sigma^m e^{-\tilde{\sigma}} 
\partial_m \tilde{\sigma} \;.
\end{equation}
The integration is performed by integrating over a three-sphere 
of radius $r$ and taking $r\rightarrow \infty$. Therefore 
the only terms in the integrand which give a finite contribute are those which 
fall off like $\frac{1}{r^3}$. The behaviour of the integrands in 
this limit is obtained by observing that 
$\hat{\cal V}(\sigma)$, and, hence, $e^{\tilde{\sigma}}$ are
algebraic functions of the harmonic functions $H_I$. Since we 
normalize the five-dimensional metric to approach the standard
Minkowski metric at infinity, both expressions approach the 
constant value 1 at infinity.
This implies the 
following  Taylor expansion of $\hat{\cal V}(\sigma)$ 
around $\tau=\frac{1}{r^2} = 0$:
\begin{eqnarray}
\hat{\cal V}(\sigma) & = & 
1 + {\cal O}(\frac{1}{r^2}) \;.\nonumber\\
\partial_m \hat{\cal V}(\sigma) 
&=& {\cal O}(\frac{1}{r^3}) \;.\nonumber 
\end{eqnarray}
This in turn implies that 
\begin{eqnarray}
e^{-\tilde{\sigma}} &=& 1 + {\cal O}(\frac{1}{r^2}) \;,\nonumber\\
\partial_m \tilde{\sigma} &=& {\cal O}(\frac{1}{r^3}) \;.\nonumber 
\end{eqnarray}
As a consequence, the factor $e^{-\tilde{\sigma}}$ in 
(\ref{ADMvsInst}) does not contribute to the integral, 
and the ADM mass and the instanton agree, 
\[
M_{ADM} = S_{\rm Inst} \;,
\]
despite
that the integrands of the surface integrals are different.\footnote{
For the special case of the dilaton-axion system, this was also
observed in \cite{EucIII}.}
This is the same result as we found when lifting without
coupling to gravity. In absence of gravity
mass is defined as the integral of the energy density, but no such 
definition is available in the presence of gravity. Instead
one needs to apply the ADM definition of mass. The fact that
we find agreement between mass and instanton action in 
both cases provides additional support for the definition 
of the instanton action obtained by dualization of axions
into tensors.

\subsection{Black hole entropy and the size of the throat}

Besides the ADM mass, the black hole entropy is the most important
property of a black hole. To extend our instanton -- black hole
dictionary we investigate the behaviour of the five-dimensional metric
at the centers and interprete it in terms of four-dimensional 
quantities. 

Line elements of the form
\[
ds^2_{(5)} = - e^{2\tilde{\sigma}} dt^2 + e^{-\tilde{\sigma}} \delta_{mn}
dx^m dx^n
\]
describe extremal black holes with the horizon located at $r=0$
if the function $e^{-\tilde{\sigma}}$ 
has the asymptotics
\[
e^{-\tilde{\sigma}} \approx \frac{Z}{r^2} \;,
\]
where $Z$ is constant. Here we use a spherical coordinate system 
which is centered at the black hole horizon.
The asymptotic line element
\[
ds^2_{(5)} = - \frac{r^4}{Z^2} dt^2 + \frac{Z}{r^2} dr^2 
+ Z d \Omega^2_{(3)} 
\]
is locally isometric to $AdS^2 \times S^3$, and the area $A$ of the event
horizon is given by the area $A=2\pi^2 Z^{3/2}$ of the
asymptotic three-sphere located at $r = 0$. 

To obtain a four-dimensional interpretation, we consider a 
hypersurface of constant time. The resulting four-dimensional
line element
\[
ds^2_{(4)} = e^{-\tilde{\sigma}} \delta_{mn}
dx^m dx^n
\]
describes the instanton in a conformal frame which is different
from the (four-dimensional) Einstein frame employed so far. 
We will call this frame the Kaluza-Klein frame, and refer to 
\cite{EucIII} for a more detailed discussion of its role and
properties.
By definition, the four-dimensional Kaluza-Klein metric is 
the pull back of the five-dimensional metric onto a hypersurface
$t=\mbox{const.}$, i.e. a constant time hypersurface of the 
black hole space-time.
In this frame the instanton line element is not flat, 
but only conformally flat. The
geometry can  
be interpreted as a semi-infinite wormhole, which is asymptotically
flat for $r\rightarrow \infty$ and ends with a neck of size
proportional to the area $A$ of the black hole for $r\rightarrow 0$. 
For multi-centered solutions there are several such throats 
with asymptotic sizes given by the areas of the corresponding horizons.

If the constant $Z$ vanishes, the area of the black hole and the
neck of the corresponding wormhole have zero size. As is well known
from supergravity solutions\footnote{See for example \cite{Mal}.}, 
a non-vanishing $Z$ requires `to switch
on sufficiently many charges'. A more precise statement will be made
later when we consider explicit examples. 
Solutions with $Z=0$ can be interpreted as degenerate
black hole solutions with vanishing area of the event horizon.
In this case the horizon coincides with the curvature singularity,
and the space-time has a null singularity. 
The spatial geometry corresponds to a 
semi-infinite wormhole with zero-sized neck. 
One expects that 
a finite horizon is obtained when taking into account higher
curvature corrections to the Einstein-Hilbert term \cite{Small}. 
Such black holes
are called small black holes, in contrast to large black holes which
already have a finite horizon at the two-derivative level.

\subsection{Attractor behaviour and examples}

We will now consider some explicit examples for illustration.
Then we return to the general case and show that the asymptotic
behaviour at the event horizons is governed by an attractor 
mechanism which generalizes the one of five-dimensional supergravity.

\subsubsection{Prepotential $\hat{\cal V}(\sigma) = \sigma^1 \sigma^2 \sigma^3$}

We start with the STU-type prepotential
$\hat{\cal V}(\sigma) = \sigma^1 \sigma^2 \sigma^3$. 
Like all models with a homogeneous cubic prepotential this 
model is supersymmetric, or more precisely, a subsector of a 
supersymmetric model \cite{EucIII}.
The dual coordinates
are $\sigma_I \simeq \partial_I \log(\sigma^1 \sigma^2 \sigma^3)
\simeq \frac{1}{\sigma^I}$,
and for convenience we fix the normalization to
\[
\sigma_I = \frac{1}{\sigma^I} \;.
\]
Then the four-dimensional instanton solution is given by
\[
\sigma^I(x) = \frac{1}{H_I(x)} \;,
\]
where $x\in \mathbbm{R}^4$. The Kaluza-Klein scalar $\tilde{\sigma}$
is 
\[
e^{3 \tilde{\sigma}} = \hat{\cal V}(\sigma) = \sigma^1 \sigma^2 \sigma^3
= \frac{1}{H_1 H_2 H_3} \;.
\]
The resulting five-dimensional line element is
\begin{eqnarray}
ds^2_{(5)} &=& - e^{2\tilde{\sigma}} dt^2 + e^{-\tilde{\sigma}} \delta_{mn}
dx^m dx^n \nonumber \\
&=& - (H_1 H_2 H_3)^{-2/3} dt^2 + 
(H_1 H_2 H_3)^{1/3} \delta_{mn} dx^m dx^n \;, \nonumber 
\end{eqnarray}
which is the standard form of a (single or multi-centered) five-dimensional 
BPS black hole for an STU-model. Observe that
the asymptotic metric at the centers is $AdS^2 \times S^3$ if
all three harmonic functions are non-constant. This requires to
have three non-vanishing charges $q_1, q_2, q_3$.
If one or two charges are switched off, one obtains `small' black holes
with vanishing horizon area. 

The result for the five-dimensional scalars $h^I$ is:
\[
h^I = e^{-\tilde{\sigma}} \sigma^I = \left( \frac{H_J H_K}{H_I^2}
\right)^{1/3} \;,
\]
where $I,J,K$ are pairwise distinct. Observe that the $h^I$ 
take finite fixed point values at the centers, which only depend
on the charges. For concreteness, single-centered harmonic
functions $H_I = h_I + \frac{q_I}{r^2}$ give
\[
\begin{CD}
h^I 
@>>{r\rightarrow 0}>  
\left( \frac{q_J q_K}{q_I^2} 
\right)^{1/3} \;.
\end{CD}
\]
A particular subclass is provided by double-extremal 
black holes, where the scalars $h^I$ are constant. 
The fixed point behaviour implies that these constant
values are not arbitrary, but determined by the charges.
For double-extremal black holes the harmonic functions $H_I$
must be proportional to one another, and the line element
takes the form
\begin{equation}
\label{Tanghelini}
ds_{(5)}^2 = - H^{-2}(x) dt^2  + H(x) \delta_{mn} dx^m dx^n \;,
\end{equation}
where $H(x)$ is a harmonic function. This is the Tanghelini
solution, which is the five-dimensional version of the extremal
Reissner-Nordstr\"om solution \cite{Tanghelini}.

Models with a general homogeneous cubic prepotential 
can be treated in an analogous way. However, in general
it is not possible to find explicit expressions for
the scalars $h^I$ or $\sigma^I$ in terms of the harmonic functions.

\subsubsection{Prepotential $\hat{\cal V}(\sigma) = \sigma^1 \sigma^2 \sigma^3
\sigma^4$ }

Let us also consider one example which does not correspond to
a supersymmetric model. We take the simplest example 
$\hat{\cal V}(\sigma) = \sigma^1 \sigma^2 \sigma^3 \sigma^4$ 
of a homogeneous quartic prepotential. We normalize the dual
scalars such that 
\[
\sigma_I = \frac{1}{\sigma^I} \;,
\] 
so that the solution is given by
\[
\sigma^I(x) = \frac{1}{H_I(x)} \;.
\]
The corresponding Kaluza-Klein scalar $\tilde{\sigma}$ is 
\[
e^{4\tilde{\sigma}} = \hat{\cal V}(\sigma) = \sigma^1 \sigma^2
\sigma^3 \sigma^4 = \frac{1}{H_1 H_2 H_3 H_4} \;,
\]
which leads to a five-dimensional line element of the form
\[
ds_{(5)}^2 = - (H_1 H_2 H_3 H_4)^{-2} dt^2 +
(H_1 H_2 H_3 H_4)^{4} \delta_{mn} dx^m dx^n \;.
\]
Multi-centered black hole solutions with finite horizons 
are thus obtained if all four harmonic functions are
non-constant, i.e. one needs four non-vanishing charges $q_1, \ldots, 
q_4$. The solution for the five-dimensional
scalars is
\[
h^I = e^{-\tilde{\sigma}} \sigma^I = \left(
\frac{H_J H_K H_L}{H_I^3} \right)^{1/4} \;,
\]
where $I, J, K, L$ are pairwise distinct. Again we observe
attractor behaviour, as the five-dimensional scalars 
approach fixed point values at the centers which only depend
on the charges. For a single-centered solution we find
\[
\begin{CD}
h^I @>>{r\rightarrow 0}> \left(
\frac{q_J q_K q_L}{q_I^3} \right)^{1/4} \;.
\end{CD}
\]
If the scalars are frozen to their fixed point values we 
obtain a double extreme solution with a Tanghelini 
line element (\ref{Tanghelini}). 

\subsubsection{General homogeneous prepotential $\hat{\cal V}(\sigma)$}

We now return to the general case and consider a  
general homogeneous prepotential. The dual coordinates
\[
\sigma_I \simeq \frac{\hat{\cal V}_I}{\hat{\cal V}}
\]
are homogenous functions of degree $-1$. The solution is
given by $\sigma_I(x) = H_I(x)$, where $H_I(x)$ are harmonic functions.
While we cannot solve this for the scalars
$\sigma^I$ in closed form, we know that the dual scalars behave like
$\sigma_I \sim \frac{1}{r^2}$ at the centers, which implies
$\sigma^I \sim r^2$. The asymptotics of the metric
at the centers is determined by 
\[
e^{-\tilde{\sigma}} = \hat{\cal V}^{-1/p} \approx  \frac{Z}{r^2} \;,
\]
and a finite event horizon requires finite $Z$. This imposes constraints
on the charges, which we discuss below.

If we express the relation
 $\sigma_I = H_I$ in terms of five-dimensional quantities we 
obtain 
\[
e^{-\tilde{\sigma}} \frac{\partial \hat{\cal V}(h)}{\partial h^I}
= H_I
\]
This has the same form as the generalized stabilisation 
equations of five-dimensional supergravity \cite{ChaSab} 
and should be
interpreted as a generalisation thereof. The generalized 
stabilisation equations are the algebraic version of the
first order flow equations which determine the black hole
solution globally. 

The stabilization or attractor equations which determine the
behaviour at the centers can be obtained by taking the limit
$r\rightarrow 0$. In this limit we have 
\[
H_I \approx \frac{q_I}{r^2} \;,\;\;\;
e^{-\tilde{\sigma}} \approx \frac{Z}{r^2} \;.
\]
The limit $r\rightarrow 0$ of the generalized 
stabilisation equation gives
\[
Z \left. \frac{\partial \hat{\cal V}(h)}{\partial h^I} \right|_*
= q_I
\]
where $*$ denotes the evaluation at the horizon. This has
the same form as the stabilisation equations (attractor equations)
of five-dimensional supergravity \cite{ChaSab}
and should be interpreted as
a generalisation thereof. 
Since
\[
\frac{\partial \hat{\cal V}(h)}{\partial h^I} h^I = 
p \hat{\cal V}(h) = p \;,
\]
the constant $Z$ can be expressed as
\[
Z = \frac{1}{p} q_I h^I_*  \;.
\]
Thus the area of the event horizon of the black hole and the
size of the neck of the corresponding wormhole/instanton
are determined by $Z$ through the charges $q_I$ 
and the attractor values of the
scalars $h^I$. For supersymmetric models ($p=3$), $Z$ is proportional
to the five-dimensional central charge.

We can be more specific about the conditions leading to 
a non-vanishing $Z$
if we restrict the functional form of 
$\hat{\cal V}(\sigma)$. 
Consider the case where $\hat{\cal V}(\sigma)$ is a homogeneous
polynomial of degree $p>0$, 
\[
\hat{\cal V}(\sigma) = C_{I_1 \cdots I_p} \sigma^{I_1} \cdots 
\sigma^{I_p} \;.
\]
Then the dual fields have the form
\[
\sigma_I \simeq \frac{\partial_I 
C_{I_1 \cdots I_p} \sigma^{I_1} \cdots 
\sigma^{I_p}}{C_{I_1 \cdots I_p} \sigma^{I_1} \cdots 
\sigma^{I_p}} \;.
\]
Two extremal situations can arise. If the prepotential has
the form 
\[
\hat{\cal V}(\sigma) = \sigma^1 \cdots \sigma^p \;,
\]
then the solution is given by 
\[
\hat{\cal V} = (H_1 \cdots H_p)^{-1}
\]
and 
\[
e^{-\tilde{\sigma}} = \hat{\cal V}(\sigma)^{-1/p}
= (H_1 \cdots H_p)^{1/p} \;.
\]
In this case a finite horzion requires that 
all charges are switched on, i.e. 
$q_1 \not=0$, \ldots $q_p \not=0$. 
The other extreme case is a prepotential of the form  
$\hat{\cal V} = \sigma^p$. 
Then the solution is given by 
\[
\hat{\cal V} = H^{-p}
\]
and
\[
e^{-\tilde{\sigma}} = \hat{\cal V}(\sigma)^{-1/p} = H \;.
\]
In this case it is sufficient to switch on one single charge, because
the corresponding scalar enters into the prepotential with the $p$-th power.
General homogeneous prepotentials provide examples for all
kinds of cases which lie between these two extreme cases.

The results of this section generalize the 
results on five-dimensional
BPS black holes to a much larger class of non-supersymmetric
models defined by homogeneous prepotentials. We observe that
for the attractor mechanism the five-dimensional scalars $h^I$
are more suitable, since they have finite fixed point values
while the four-dimensional scalars $\sigma^I$ go to zero.
However, since $h^I$ and $\sigma^I$ are related by a 
rescaling, both descriptions are equivalent, and the 
asymptotic fixed point of infinity of the $\sigma^I$ corresponds
to the proper fixed point for the $h^I$.

\section{Conclusions and Outlook}

In this paper we have constructed multi-centered
extremal black hole solutions using temporal
reduction without imposing spherical symmetry.
By imposing that the solution can be expressed algebraically in 
terms of harmonic functions, we have identified a class 
of scalar geometries which is characterized by the 
existence of (Hesse or para-K\"ahler) potential for the metric. 
This class of theories contains supergravity theories as 
a subset while preserving the salient feature of 
BPS solutions, namely multi-centered generalizations and
the generalized stabilisation equations. 
The distinction between BPS and
non-BPS extremal solutions in supergravity has been subsumed
under the geometrical distinction between solutions which flow along
eigendistributions of the para-complex structure and 
those which flow along other completely isotropic submanifolds
of the (extended) scalar manifold. 
Starting from
the interpretation of the equation of motion as
defining a harmonic map between the (reduced) space-time
and the (extended) scalar manifold, the solution can
be expressed algebraically in terms of harmonic functions
without the need to bring the equations of motion to
first order form. A first order rewriting can still be
obtained by imposing that the solution carries finite 
charges. We plan to use this link to explore the relation
between our formalism and the approaches using first order
rewritings, `fake'-supersymmetry and Hamilton-Jacobi theory.
It should also be interesting to investigate how Hessian
scalar manifolds could be used within the entropy function
formalism of Sen \cite{SenEntropyFunction}. This approach 
allows to study the near horizon geometry of generic
Einstein-Maxwell type theories, but it is in general not
possible to learn much about the extension of solutions
away from the horizon. For BPS solutions one can make
the transition from near-horizon to global solutions
because generalized stabilisation equations and the
`proper' stabilisation equations have the same structure,
and we have seen that this feature generalizes to a
large class of non-supersymmetric theories. The electric 
BPS-solutions of five-dimensional are a subclass of
our solutions, and one expects that the corresponding
instantons are BPS solutions of the reduced four-dimensional
Euclidean theory. This can indeed be verified directly,
and in \cite{MohWai1} we will give a more detailed 
account on instanton solutions for Euclidean vector 
multiplets.

For concreteness, we have restricted ourselves in this paper to the
relation between five-dimensional Einstein-Maxwell theories
and four-dimensional Euclidean sigma models, and to extremal and
electro-static backgrounds. This leaves various 
directions for future work. Evidently, many features
of our constructions will generalize to any number of
dimensions, the most interesting pair being four-dimensional
Einstein-Maxwell type theories and three-dimensional sigma models.
Moreover, there are various other types of solutions, like
black holes in anti-de-Sitter and de-Sitter space, rotating
black holes, black strings and black rings, Taub-NUT spaces, 
solutions including higher curvature terms, and non-extremal
solutions. While some of these 
might just correspond to more complicated harmonic maps, others
will require generalizations of the set up, since the 
temporal reduction will in general lead to Euclidean theories 
which also contain gauge fields and a scalar potential.
It will be interesting to see whether solutions can be
constructed efficiently in such a generalized set up.
In this respect it is encouraging that black ring 
solutions for five-dimensional 
Einstein-Maxwell-Dilaton gravity have been
constructed by lifting solutions of four-dimensional
Euclidean sigma models with (symmetric) para-complex
target spaces \cite{Yazadjiev}.

Besides the construction and study of solutions, the geometrical
structures underlying the Lagrangians are very interesting.
Both \cite{AleCor} and our work suggest that there are 
natural generalizations of the special geometries
realized in supersymmetric theories. Besides the existence
of a potential,  homogeneity conditions play an important
role, which indicates that the underlying manifolds have
homothetic Killing vector fields. This is well known
feature of the scalar geometries of 
vector, tensor and hypermultiplets when these are considered
in the superconformal formalism. 

As mentioned at various places in this paper, the scalar 
geometries of theories obtained from the same higher-dimensional
theory by dimensional reduction over space and time, respectively,
are related by analytic continuation. We have also noticed
that this is related to an ambiguity in singling out `the'
Euclidean action of a given theory. This has been discussed 
in some detail in \cite{EucIII}, the role of these ambiguities
for instanton effects is currently under investigation \cite{MohWai}.
Here we would like to point out that these ambiguities 
suggest to work
in the framework of complex-Riemannian geometry and to 
regard scalar manifolds which are related by analytic continuation
as real forms of a single underlying 
manifold. Interestingly, similar analytic continuations,
the complexification of field space, and the subsequent 
classification of reality conditions seesm to play an important
role in recent studies of 
black holes, instantons, domain walls and cosmological solutions
within the framework of `fake'-supersymmetry, see for example 
\cite{Fake2,Bergshoeff:Complex}. While so far such investigations 
have been restricted to symmetric
target spaces, complex-Riemannian geometry should provide
the appropriate framework for extening these studies to 
general targtes. Some elements needed for this are 
provided in the appendix.

\begin{appendix}
\section{Complexification of the target space}

At the end of Section \ref{SectEucAct} we observed that
the target spaces $M$ and $M'$ of the 
two Euclidean actions (\ref{paraH-real}) and 
(\ref{Herm-real}) (equivalently (\ref{SX}) an (\ref{SY}))
can be viewed as real sections of one underlying complex
manifold. Complexification of the action is used in 
some approaches to defining the Euclidean actions of
supersymmetric theories \cite{vNWal}. 
Complex actions for the ten-dimensional and eleven-dimensional
supergravity theories were discussed in \cite{Bergshoeff:Complex},
while \cite{EucIII} found that a similar formalism should be
useful for Euclidean vector multiplets in four dimensions.

Since the 
scalar target spaces of (\ref{Herm-real}) and (\ref{paraH-real})
already a complex or para-complex
structure, respectively,  before we complexify them
some care is needed in order to distinguish between 
the different complex structures.\footnote{If one includes fermions
then yet another complex structure becomes relevant, namely the one carried
by the spinor representation \cite{EucI}. Here we will restrict ourselves to 
bosonic actions. } In the following we work out some details
and arrive at the conclusion that the proper geometrical 
framework for complexified Euclidean actions is complex-Riemannian
geometry. 

When we use the real coordinates $(\sigma^I, b^I)$, then the
metrics of the target spaces $M$ and $M'$, which underly the 
actions (\ref{paraH-real}) and (\ref{Herm-real}) 
have the form
\[
ds^2 = N_{IJ}(\sigma) ( d\sigma^I d\sigma^J \mp db^I db^J) \;,
\]
respectively. The two line elements
are related by the analytic continuation $b^I \rightarrow i b^I$.
If we complexify the $b^I$, then $M$ and $M'$ can
be viewed as subspaces of a $3n$-dimensional space $\tilde{M}$.

This description is not satisfactory for various reasons.
Complexifying only the $b^I$ introduces an asymmetry between
the $\sigma^I$ and the $b^I$. It is more natural to complexify
all fields and to view $M$ and $M'$ as real forms of a complex
space $M_c$. Moreover, $M$ and $M'$ carry additional structures.
$M$ is a complex space, and when we use
complex coordinates $Y^I = \sigma^I + i b^I$ the line element
of $M'$ is manifestly Hermitian
\[
ds^2_{M'} = N_{IJ}(Y+\bar{Y}) dY^I d\bar{Y}^J \;.
\]
If we want to view $M'$ as a real form of a complex space
$M_c$, then we need to be careful in distinguishing the complex
structure of $M'$, and the complex structure of $M_c$ which
is introduces in the process of complexification, and which is 
used in
the analytic continuation from $M'$ to $M$. Similarly $M$
is a para-complex space and when
using the para-complex coordinates $X^I = \sigma^I + e b^I$,
the line element of $M$ is manifestly para-Hermitian:
\[
ds^2_{M} =  N_{IJ}(X+\bar{X}) dX^I d\bar{X}^J \;.
\]

In the following\footnote{The mathematical background material
relevant for the following paragraphs can be found in \cite{Lou,Spin}
and \cite{EucI,EucIII}.}
we will reserve the symbol $i$ for the
imaginary unit associated with the complex structure 
of $M$, while the imaginary unit associated with the complex
structure of $M_c$ will be denoted $j$. We can define
$j$ in terms of $i$ and $e$ by observing that
the analytic continuation from $M'$ to $M$, when written
in para-complex coordinates, takes the form
\[
Y^I = \sigma^I + i b^I \rightarrow 
X^I = \sigma^I + e b^I \;.
\]
The replacement $ib^I \rightarrow eb^I$ is induced 
by $b^I \rightarrow (-ie) b^I$, and implies that $j$ 
should be defined as $j=-ie$. To have $j^2=-1$ we need
to impose the relation $ie=ei$. These relations are consistent
and define a four-dimensional commutative and assosiative
real algebra, with basis $1,i,e,j$. 
This 
algebra is generated by $i$ and $e$, subject to the relations
\begin{equation}
\label{Rel1}
i^2 =-1 \;,\;\;\;e^2=1\;,\;\;\;ie=ei \;,
\end{equation}
and defining $j=-ie$. 
Equivalently, this algebra 
is generated by $i$ and $j$ 
subject to the relations
\[
i^2 = -1 \;,\;\;\;j^2=-1 \;,\;\;\;ij=ji \;\;\;\;
\]
and defining $e=ij$.The second presentation shows that the
algebra is isomorphic to $\mathbbm{C} \oplus \mathbbm{C}$. 
Note that this is not only an algebra over $\mathbbm{R}$ 
but also an algebra over $\mathbbm{C}$.

To see that the complex algebra $\mathbbm{C} \oplus \mathbbm{C}$
is the `complexification of the complex
numbers' $\mathbbm{C}$ as well as the 
`complexification of the para-complex numbers' $C$, recall
that  the complexification of a real algebra $A$ (associative,
with unit) is obtained by taking the real tensor product
(of algebras) with $\mathbbm{C}$ (considered as a real algebra):
\[
A_c = \mathbbm{C} \otimes_{\mathbbm{R}} A \;.
\]
If one takes $A=\mathbbm{C}$ then the result is 
\begin{equation}
\label{ComplexifyComplex}
\mathbbm{C} \otimes_{\mathbbm {R}} \mathbbm{C} \simeq \mathbbm{C} 
\oplus \mathbbm{C} \;.
\end{equation}
If one takes $A$ to be the para-complex numbers 
$C$, one obtains the same result:
\begin{equation}
\label{ComplexifyParaComplex}
\mathbbm{C} \otimes_{\mathbbm{R}} C \simeq \mathbbm{C} \oplus
\mathbbm{C}
\end{equation}
The isomorphisms  (\ref{ComplexifyComplex}) and (\ref{ComplexifyParaComplex})
can easily be written down explicitly in terms of bases. 
Alternatively we can simply note
that $\mathbbm{C}$ and $C$ are real Clifford algebras:
\[
Cl_{1,0} \simeq \mathbbm{C} \;,\;\;\;
Cl_{0,1} \simeq \mathbbm{R} \oplus \mathbbm{R} \simeq C \;,
\]
which is obvious from the relations $i^2=-1$ and $e^2=1$ of the
generating elements $i$ and $e$. Given this, we can refer to the 
known fact 
that the complexifications of these two Clifford algebras 
are \cite{Lou,Spin}:
\[
\mathbbm{C}l_1 = 
\mathbbm{C} \otimes_{\mathbbm{R}} Cl_{1,0} =
\mathbbm{C} \otimes_{\mathbbm{R}} Cl_{0,1} \simeq
\mathbbm{C} \oplus \mathbbm{C} \;.
\]

For models with $2n$ free scalar fields the target spaces
are simply (the affine spaces associated to the) 
vector spaces $M=C^n$ and $M'=\mathbbm{C}^n$.
Both are real $2n$-dimensional vector spaces, but carry
additional structures: $M'=\mathbbm{C}^n$ is a (complex-)
$n$-dimensional vector space, $M=C^n$ is an $n$-dimensional
free module over the algebra of para-complex numbers
$C$ \cite{EucI,EucIII}. Since the complexifications of
the underlying algebras $C$ and $\mathbbm{C}$ coincide,
the complexifications of $M$ and $M'$ are also isomorphic:
\[
M_c \simeq \mathbbm{C} \otimes_{\mathbbm{R}} \mathbbm{C}^n 
\simeq \mathbbm{C} \otimes_{\mathbbm{R}} C^n \simeq
\mathbbm{C}^n \oplus \mathbbm{C}^n \simeq \mathbbm{C}^{2n} \;.
\]
For models with $2n$ interacting fields, the target spaces 
$M$ and $M'$ are (para-)complex
manifolds, i.e. manifolds modelled on $C^n$ and $\mathbbm{C}^n$, respectively.
Both are in particular $2n$-dimensional real manifolds, 
and the complexification 
$M_c$ is a (complex-)$2n$-dimensional complex manifold. 
The tanget spaces of $M,M',M_c$ are $T_PM=C^n$, $T_PM'=\mathbbm{C}^n$
and $T_P M_c = \mathbbm{C}^{2n} \simeq \mathbbm{C}\otimes_{\mathbbm R}
C^n \simeq \mathbbm{C} \otimes_{\mathbbm{R}} \mathbbm{C}^n$, 
respectively.

The dynamics
of the scalar fields is controlled by the (pseudo-)Riemannian 
metrics of $M$ and $M'$. To study the effect of complexification,
let us start with the case of 2 free real scalar fields $\sigma$ and
$b$. Then the real, positive definite line element of $M'$ is
\[
ds^2_{M'} = d \sigma d \sigma + d b db \;.
\]
This can be complexified by promoting the real fields $\sigma, b$ 
to complex fields:
\begin{eqnarray}
\sigma &\rightarrow & \Sigma = \sigma_1 + j \sigma_2 \;, \nonumber \\
b & \rightarrow & B = b_1 + j b_2 \;.\nonumber  \label{Complexification}
\end{eqnarray}
Here $j$ is the imaginary unit associated with the complex structure
of $M_c$. The resulting complex line element is
\[
ds^2_{M'_c} = d \Sigma d \Sigma + d B d B = 
[d \sigma_1 d \sigma_1 - d\sigma_2 d \sigma_2  
+ db_1 db_1 - d b_2 db_2] 
+ 2j [d \sigma_1 d\sigma_2 
+ d b_1 db_2] \;.
\]
The line element of $M'$ is recovered by taking the real section 
$\sigma_2 = b_2=0$. If we take instead the real section 
$\sigma_2=b_1=0$ we obtain the real line element
\[
ds^2 = d \sigma_1 d \sigma_1  - d b_2 db_2 \;,
\]
which has split signature. Upon setting $\sigma_1=\sigma$
and $b=b_2$ we obtain the line element 
\[
ds^2_M = d\sigma d\sigma - d b db
\]
of $M$. Conversely, if we complexify $ds^2_M$ by (\ref{Complexification}),
then we obtain complex line element
\[
ds^2_{M_c} = d \Sigma d\Sigma - d B d B =
[d \sigma_1 d \sigma_1 - d\sigma_2 d \sigma_2  - db_1 db_1 + d b_2 db_2] 
+ 2j[ 
d \sigma_1 d\sigma_2 -  d b_1 db_2] \;.
\]
The line elements $ds^2_{M_c}$ and $ds^2_{M'_c}$ are related
by the $B \rightarrow j B$, and define the same complex metric on 
$M_c$. 

The complexification can also be formulated in terms of the
complex field $Y=\sigma + ib$. Here the distinction between
$i$ and $j$ is important to avoid confusion.
 In terms of $Y$, the line element
of $M$ is
\[
ds^2_M= dY d\bar{Y} \;.
\]
Here `complexification of $Y$' can be understood as `taking $Y$ and
$\bar{Y}$ to be independent complex variables'. Using the 
distinction between $i$ and $j$ we can make this precise:
\begin{eqnarray}
Y = \sigma + i b &\rightarrow& \Sigma + i B \;,\nonumber \\
\bar{Y} = \sigma - i b &\rightarrow& \Sigma - i B \;,\nonumber 
\end{eqnarray}
with complex fields $\Sigma = \sigma_1 + j \sigma_1$ and 
$B= b_1 + j b_2$.  Similarly, the complex line element of $M_c$
can be obtained by `complexifying the para-complex field' 
$X=\sigma +e b$.

The most general case we are interested in are line elements of 
the form
\[
ds^2_{M/M'} = N_{IJ}(\sigma) (d\sigma^I d \sigma^J \pm db^I d b^J) \;.
\]
The increase in the number of fields does not change much, 
as we only need to introduce indices $I,J$ to label the fields.
If the real metric  $N_{IJ}(\sigma)$ is not flat, we need to assume that 
it is real-analytic in the $\sigma^I$, so that it can be extended
analytically to a holomorphic matrix function 
$N_{IJ}(\Sigma)$,  in some neighbourhood of $\sigma^I_{2}=0$. 
The resulting complex manifold $M_c$ contains
$M$ and $M'$ as the real submanifolds $\sigma_2=b_2=0$ and
$\sigma_2=b_1=0$, respectively.
For the purpose of embedding
$M$ and $M'$ into some complex manifold, it is not relevant how we
choose the neighbourhood of $\sigma_2=0$. 
The resulting line element 
\[
ds^2_{M_c} = N_{IJ}(\Sigma) (d \Sigma^I d \Sigma^J + d B^I 
dB^J)
\]
defines a complex-Riemannian
metric on $M_c$. A complex-Riemannian metric on a complex manifold
is a complex bilinear form on the holomorphic tangent bundle.\footnote{
A definition of complex-Riemannian manifolds and some further 
references can be found in \cite{Ivanov}. 
The extension of the bilinear form 
to the anti-holomorphic tangent bundle is given
by complex conjugation. Taking the holomorphic and anti-holomorphic
tangent bundles to be orthogonal, one obtains a natural extension to the
full (complexified) tangent bundle.}
Note that K\"ahler (more generally Hermitean)
manifolds are Riemannian
manifolds and not complex-Riemannian manifolds: they carry 
a positive definite hermitean sesquilinear form on their
(complexified) tangent bundle, whose real part is a Riemannian
metric. Similarly para-K\"ahler (and pseudo-K\"ahler, para-Hermitean,
pseudo-Hermitean) manifolds
are pseudo-Riemannian, not complex-Riemannian. 
The $n$ real shift isometries $b^I \rightarrow b^I + c^I$ 
induce $n$ complex shift isometries $B^I \rightarrow B^I + C^I$
on $M_c$.  

Symmetric spaces (which are listed in \cite{Gilmore}
and \cite{Helgason})
provide plenty of examples for triples $(M,M',M_c)$.
The simplest example is
\[
M \simeq \frac{SL_2(\mathbbm{R})}{SO(1,1)}\;,\;\;\;
M' \simeq \frac{SL_2(\mathbbm{R})}{SO(2)}\;,\;\;\;
M_c \simeq \frac{SL_2(\mathbbm{C})}{GL(1,\mathbbm{C})}\;.
\]
The space $\frac{SL(2,\mathbbm{R})}{SO(1,1)}$ 
occured in several examples in the main paper. 
In Section 4 we have seen explicitly that this pseudo-Riemannian symmetric 
is para-K\"ahler, and that it is related by analytic continuation 
to  the Riemannian symmetric space
$\frac{SL(2,\mathbbm{R})}{SO(2)}$, which is K\"ahler.
Note that while the above example uses symmetric spaces, the
discussion in this appendix applies to analytic (pseudo-)Riemannian
manifolds in general.

\end{appendix}

\subsubsection*{Acknowledgements}

We would like to thank
Gabriel Lopes Cardoso and Vicente Cort\'es for various valuable
discussion. T.M. thanks the Center for Mathematical Physics
of the University of Hamburg for support and hospitality during
several visits to Hamburg. He also thanks the LMU Munich for
hospitality and the Royal Society for support of this work through
the Joint Projects Grant `Black Holes, Instantons and String 
Vacua'.

\end{document}